%% file: main.tex
\documentclass[headings=standardclasses]{scrartcl}

\usepackage{geometry}
\usepackage[utf8]{inputenc}
\usepackage[T1]{fontenc}
\usepackage[onehalfspacing]{setspace}
\usepackage[charter]{mathdesign}
\usepackage{microtype}
\usepackage{amsmath}
\usepackage{amsthm}
\usepackage{tabulary}
\usepackage{longtable}
\usepackage{booktabs} 
\usepackage{diagbox}
\usepackage{ragged2e}

\usepackage{siunitx}
\usepackage{bm}
\usepackage[pdftex]{graphicx}
\usepackage{rotating}
\usepackage{pdfpages}
\usepackage{pdflscape}
\usepackage{color}
\usepackage{array}
\usepackage[font=bf]{caption}
\usepackage{subcaption}
\usepackage{pgfplots}
\usepackage{pgfplotstable}
\usepackage{float}
\usepackage[title]{appendix}
\pgfplotsset{compat=1.9}
\usepackage{url}
\usepackage[section]{placeins}
\usepackage{natbib}
\usepackage{hyperref}
\hypersetup{
	colorlinks   = true,    
	urlcolor     = blue,    
	linkcolor    = blue,    
	citecolor    = blue      
}
\usepackage{algorithm}
\usepackage{algpseudocode}

\makeatletter
\renewcommand*\env@matrix[1][c]{\hskip -\arraycolsep
  \let\@ifnextchar\new@ifnextchar
  \array{*\c@MaxMatrixCols #1}}
\makeatother

\usepackage{xcolor}
\usepackage{soul}

\usepackage{verbatim}

\begin{document}
\thispagestyle{empty}
\begin{spacing}{1.2}
\begin{center}
\huge \textbf{Robust discrete choice models \\ with t-distributed kernel errors
} \\
\vspace{\baselineskip}
\end{center}
\begin{flushleft}
\normalsize
17 September 2022 \\
\vspace{\baselineskip}
\textsc{Rico Krueger} (corresponding author) \\
Department of Technology, Management and Economics \\
Technical University of Denmark (DTU), Denmark \\
rickr@dtu.dk \\
\vspace{\baselineskip}
\textsc{Michel Bierlaire} \\
Transport and Mobility Laboratory \\
Ecole Polytechnique F\'{e}d\'{e}rale de Lausanne, Switzerland \\
michel.bierlaire@epfl.ch\\
\vspace{\baselineskip}
\textsc{Thomas Gasos} \\
Transport and Mobility Laboratory \\
Ecole Polytechnique F\'{e}d\'{e}rale de Lausanne, Switzerland \\
thomas.gasos@alumni.epfl.ch\\
\vspace{\baselineskip}
\textsc{Prateek Bansal} \\
Department of Civil and Environmental Engineering\\
National University of Singapore, Singapore \\
prateekb@nus.edu.sg  \\
\end{flushleft}
\end{spacing}

\newpage
\thispagestyle{empty}
\section*{Abstract}

Outliers in discrete choice response data may result from misclassification and misreporting of the response variable and from choice behaviour that is inconsistent with modelling assumptions (e.g. random utility maximisation). In the presence of outliers, standard discrete choice models produce biased estimates and suffer from compromised predictive accuracy. Robust statistical models are less sensitive to outliers than standard non-robust models. This paper analyses two robust alternatives to the multinomial probit (MNP) model. The two models are robit models whose kernel error distributions are heavy-tailed t-distributions to moderate the influence of outliers. The first model is the multinomial robit (MNR) model, in which a generic degrees of freedom parameter controls the heavy-tailedness of the kernel error distribution. The second model, the generalised multinomial robit (Gen-MNR) model, is more flexible than MNR, as it allows for distinct heavy-tailedness in each dimension of the kernel error distribution. For both models, we derive Gibbs samplers for posterior inference. In a simulation study, we illustrate the excellent finite sample properties of the proposed Bayes estimators and show that MNR and Gen-MNR produce more accurate estimates if the choice data contain outliers through the lens of the non-robust MNP model. In a case study on transport mode choice behaviour, MNR and Gen-MNR outperform MNP by substantial margins in terms of in-sample fit and out-of-sample predictive accuracy. The case study also highlights differences in elasticity estimates across models.   \\
\\
\emph{Keywords:} robustness, probit, robit, Bayesian estimation, discrete choice, outliers

\newpage
\pagenumbering{arabic}

\section{Introduction}

Discrete choice models are widely used to explain and predict the demand for discrete alternatives such as retail products, medical treatments and modes of transport. Discrete choice models are usually formulated based on random utility theory \citep{mcfadden1981econometric} which posits that a rational decision-maker chooses the alternative with the highest utility from a set of mutually exclusive alternatives. In principle, utility is not identifiable in absolute terms, and only differences in utility matter. Utility differences are typically assumed to follow independent and identical logistic or multivariate Gaussian distributions. Whereas the former assumption leads to a logit kernel, the latter assumption leads to a probit kernel. 

Logit is popular in practice due to its closed-form choice probabilities. However, logit suffers from two drawbacks. First, the independence of irrelevant alternatives property implies restrictive substitution patterns. Second, logit assumes that the random error terms are homoskedastic across alternatives, which implies that the same level of decision uncertainty applies to all alternatives in a choice set. Probit overcomes the limitations of logit by allowing for the estimation of a full error covariance, subject to identification restrictions \citep{train2009discrete}. Probit choice probabilities lack closed-form expressions. However, advances in computational power combined with progress in simulation techniques \citep{burgette2012trace, hajivassiliou1996simulation, imai2005bayesian, mcculloch1994exact, train2009discrete} and analytical approximation approaches \citep{bhat2011maximum} have attenuated the relevance of this drawback. 

Both logit and probit lack robustness to outliers in the response data due to their light-tailed kernel error distributions \citep{benoit2016outlier, hausman1998misclassification}. Outliers in discrete choice response data may result from various contamination processes, including misclassification and misreporting of the response variable as well as choice behaviour that is inconsistent with the postulated decision rule (e.g. random utility maximisation). Ignoring outliers in logit and probit applications can lead to biased and inconsistent parameter estimates \citep{hausman1998misclassification}. 
Even though response data contamination is a well acknowledged problem in discrete choice analysis \citep{hausman1998misclassification, paleti2019misclassification}, the development of robust multinomial choice models is understudied. Extant approaches lack flexibility and suffer from computational drawbacks. 

In this paper, we address this research gap by analysing two robust alternatives to the multinomial probit (MNP) model. 
Both models belong to the family of robit models whose kernel error distributions are heavy-tailed t-distributions to moderate the influence of outliers. 
In the first model, the multinomial robit (MNR) model, a generic degrees of freedom parameter controls the heavy-tailedness of the kernel error distribution. The second model, the generalised multinomial robit (Gen-MNR), is more flexible than MNR, as it allows for different marginal heavy-tailedness of the kernel error distribution. 
For both models, we devise Gibbs samplers for posterior inference.
We first use simulated data to investigate the properties of the proposed models and their estimation methods in terms of parameter recovery and elasticity estimates. 
Then, we compare MNP, MNR and Gen-MNR in a case study on transport mode choice behaviour in London, UK.

The remainder of this paper is structured as follows: 
In Section~\ref{sec:related}, we elaborate on the research gap and discuss connections to related work.
In Section~\ref{sec:form}, we exhibit the mathematical formulations of MNP, MNR and Gen-MNR.
In Section~\ref{sec:infer}, we describe inference and implementation details. 
In Section~\ref{sec:sim}, we present the simulated examples; in Section~\ref{sec:case}, we present the case study. 
Finally, we conclude in Section~\ref{sec:conc}.

\section{Related work} \label{sec:related}

Robust statistical models are less sensitive to outliers than standard ``non-robust'' models \citep{gelman2013bayesian}. Heavy-tailed distributions such as the t-distributions are an appealing choice for enhancing the robustness of statistical models \citep{gelman2006data, lange1989robust}. Compared to the Gaussian distribution, the t-distribution has one more parameter, called the degrees of freedom (DOF) parameter, which controls the heavy-tailedness of the distribution and moderates outliers. 

There are several pertinent examples of outlier robust models that are based on t-distributions. In the context of generalised linear models, \citet{liu2004robit} proposes the binary robit model, which is based on a t-distribution with unknown degrees of freedom, as a robust alternative to logistic and probit regression models. Furthermore, \citet{ding2014bayesian} constructs a robust Heckman selection model using the t-distribution as kernel error distribution. \citet{jiang2016robust} formulate Heckman selection and multivariate robit models based on t-distributions with different marginal DOF. However, robust multinomial choice models with heavy-tailed kernel error distributions have received limited attention. 

Researchers have proposed various departures from standard kernel error distributions for multinomial choice models \citep[also see][for a review]{paleti2019discrete}, including generalised extreme value \citep{mcfadden1978modeling}, heteroskedastic extreme value \citep{bhat1995heteroscedastic}, negative exponential \citep{alptekinouglu2016exponomial, daganzo1979multinomial}, negative Weibull \citep{castillo2008closed}, generalised exponential \citep{fosgerau2009discrete} and q-generalised reverse Gumbel \citep{chikaraishi2016discrete} kernel error distributions, additive combinations of Gumbel and exponential error terms \citep{del2016class}, a class of asymmetric distributions \citep{brathwaite2018asymmetric} and copulas with Gumbel marginals \citep{del2020choice}. 
However, none of these advancements aim at enhancing robustness while also admitting flexible substitution patterns and heteroskedasticity.

To the best of our knowledge, only three studies formulate robust multinomial choice models using heavy-tailed distributions. 
\citet{dubey2020generalized} present the first multinomial robit (MNR) model, i.e. a multinomial choice model defined through a t-distributed kernel error with an estimable DOF. \citet{dubey2020generalized} make a strong empirical case to adopt the MNR model over the multinomial probit (MNP) model. First, the estimates of the MNP model are inconsistent, if the kernel errors in the data generating process are heavy-tailed. Second, the robustness of MNR results in superior in-sample fit and out-of-sample predictive ability for class-imbalanced datasets.
Furthermore, \citet{peyhardi2020robustness} formulates a MNR model in the context of generalised linear models and shows that MNR can help in identifying artificial aspects in the design of stated preference experiments. 
Finally, \citet{benoit2016outlier} devise a multinomial choice model, in which utility differences follow a symmetric and heavy-tailed multivariate Laplace distribution. Using simulated and real data, the authors show that their proposed model is less sensitive to outliers than MNP.

We identify two research gaps in the formulation and estimation of existing robust multinomial choice models.
First, the kernel error distributions of existing models lack flexibility.
\citet{dubey2020generalized} and \citet{peyhardi2020robustness} constrain the flexibility of the kernel error distribution by assuming that a single, generic DOF parameter controls the heavy-tailedness of the kernel error distribution. This modelling assumption implies that all utility differences between different pairs of alternatives exhibit the same level of aberrance. 
Unlike the t-distribution, the Laplace distribution underlying the formulation of the multinomial choice model proposed by \citet{benoit2016outlier} cannot exhibit varying levels of heavy-tailedness, since the Laplace distribution does not have a third parameter controlling the heavy-tailedness of the distribution.
Second, the estimation approaches employed in the studies by \citet{dubey2020generalized} and \citet{peyhardi2020robustness} are not scalable. \citet{dubey2020generalized} are unable to derive analytical gradients of the MNR model and thus rely on computationally-expensive numerical gradient approximations during the maximisation of the simulated log-likelihood of the model. \citet{peyhardi2020robustness} estimates the DOF parameter by performing a grid search, which requires the model to be estimated at multiple values of the DOF parameter. Clearly, this approach suffers from the curse of dimensionality if the underlying kernel distribution has multiple DOF parameters. 
Importantly, in both approaches, hierarchical model extensions such as random parameters, latent variables and shrinkage priors are computational expensive, since these extensions necessitate an additional layer of simulation in the computation of the log-likelihood.

In this paper, we address the first limitation of existing MNR models (i.e. generic heavy-tailedness) by formulating a generalised MNR (Gen-MNR) model with alternative-specific DOF parameters. To that end, we adopt the non-elliptical contoured t-distribution \citep{jiang2016robust} as kernel error distribution. 
To address the second limitation (i.e. computationally-expensive estimation), we devise gradient-free Bayesian estimation approaches for both the MNR and the Gen-MNR models. In the construction of the Bayesian estimation approaches, we exploit the hierarchical normal mixture representation of the t-distribution. To circumvent complex likelihood computations in the estimation of the MNR and the Gen-MNR models, we employ a combination of Bayesian data augmentation techniques used in the estimation of MNP models \citep{albert1993bayesian,mcculloch1994exact} as well as of non-multinomial robit models \citep{ding2014bayesian,jiang2016robust}. Bayesian estimation also facilitates potential hierarchical model extensions. 

\section{Models} \label{sec:form}


\subsection{Multinomial probit (MNP)} \label{sec:model_mnp}

We consider a standard random utility model in which an agent $i = 1, \ldots, N$ chooses from a set of $J$ mutually exclusive alternatives. In principle, utility is not identified at an absolute level. Therefore, the MNP model is defined through a $J-1$-dimensional Gaussian latent variable vector $\boldsymbol{w}_{i} = \{ w_{ij}, \ldots, w_{i,J-1} \}$ \citep{mcculloch1994exact}. The elements of $\boldsymbol{w}_{i}$ correspond to the utility differences with respect to the base alternative $J$. The observed choice $y_{i} \in \{1, \ldots, J \}$ is assumed to arise from 
\begin{equation} \label{eq:lik}
y_{i}(\boldsymbol{w}_{i}) = 
\begin{cases}
j & \text{if } \max (\boldsymbol{w}_{i}) = w_{ij} > 0 \\
J & \text{if } \max (\boldsymbol{w}_{i}) < 0,
\end{cases}
\quad \text{for } i = 1, \ldots, N.
\end{equation}
The latent variable  $\boldsymbol{w}_{i}$ is represented as 
\begin{equation} \label{eq:lv}
\boldsymbol{w}_{i} = \boldsymbol{X}_{i} \boldsymbol{\beta} + \boldsymbol{\varepsilon}_{i}
\quad
\text{with } \boldsymbol{\varepsilon}_{i} \sim N(\boldsymbol{0}, \boldsymbol{\Sigma}),
\quad \text{for } i = 1, \ldots, N.
\end{equation}
Here, $\boldsymbol{X}_{i}$ is a $(J - 1) \times K$ matrix of differenced predictors, i.e. 
$\boldsymbol{X}_{i} = \begin{bmatrix} \boldsymbol{X}_{i1} \\ \vdots \\ \boldsymbol{X}_{i,J-1} \end{bmatrix}  = \begin{bmatrix} \boldsymbol{X}_{i1}^{\text{obs}} - \boldsymbol{X}_{iJ}^{\text{obs}} \\ \vdots \\ \boldsymbol{X}_{i,J-1}^{\text{obs}} - \boldsymbol{X}_{iJ}^{\text{obs}} \end{bmatrix}$, where $\boldsymbol{X}_{ij}^{\text{obs}}$ is the observed attribute vector of alternative $j$ for agent $i$. $\boldsymbol{\beta}$ is a $K$ vector of taste parameters. $\boldsymbol{\Sigma}$ is a $(J - 1) \times (J - 1)$ covariance matrix. 
The latent variable representation (\ref{eq:lv}) is not identified, because $\boldsymbol{w}_{i}$ can be multiplied by any positive scalar $c$ without changing the likelihood (\ref{eq:lik}), i.e. $y_{i}(\boldsymbol{w}_{i}) = y_{i}(c \boldsymbol{w}_{i})$.
Therefore, we must set the scale of the model. 
Following \citet{burgette2012trace}, we impose a trace restriction on $\boldsymbol{\Sigma}$ with $\text{tr}(\boldsymbol{\Sigma}) = J - 1$. 
We implement the trace restriction using a constrained inverse Wishart prior \citep{imai2005bayesian, burgette2012trace}.
To construct this prior, we introduce a working parameter $\alpha$, which is not identified given the observed data $\boldsymbol{y}$, but is identified given $\{ \boldsymbol{y}, \boldsymbol{w} \}$. 
We transform $\tilde{\boldsymbol{w}}_{i} = \alpha \boldsymbol{w}_{i}$ for $i = 1, \ldots, N$ so that $\tilde{\boldsymbol{w}}_{i} \sim N( \boldsymbol{X}_{i} \tilde{\boldsymbol{\beta}}, \tilde{\boldsymbol{\Sigma}})$ with $\tilde{\boldsymbol{\beta}} = \alpha \boldsymbol{\beta}$ and $\tilde{\boldsymbol{\Sigma}} = \alpha^{2} \boldsymbol{\Sigma}$. 
We then put an inverse Wishart prior on the intermediate, unidentified quantity $\tilde{\boldsymbol{\Sigma}}$ so that $\tilde{\boldsymbol{\Sigma}} \sim IW (\rho, \tilde{\boldsymbol{\Lambda}})$.
After the transformation $\boldsymbol{\Sigma} = \tilde{\boldsymbol{\Sigma}} / \alpha^{2} $ with $\alpha^{2} = \text{tr}( \tilde{\boldsymbol{\Sigma}} ) / (J - 1)$, the implied prior on the tuple $\{ \boldsymbol{\Sigma}, \alpha^{2} \}$ is 
\begin{equation} \label{eq:prior_Sigma_alpha2}
P(\boldsymbol{\Sigma}, \alpha^{2}) \propto \vert \boldsymbol{\Lambda} \vert^{- (\rho + J) / 2} e^{- \frac{\alpha_{0}^{2}}{2 \alpha^{2}} \text{tr} ( \boldsymbol{\Lambda} \boldsymbol{\Sigma}^{-1}) } (\alpha^{2})^{-\rho (J - 1) / 2 + 1} \textbf{1} \{ \text{tr}(\boldsymbol{\Sigma}) = J - 1 \},
\end{equation}
where $\alpha_{0}$ is some positive constant such that $\tilde{\boldsymbol{\Lambda}} = \alpha_{0} \boldsymbol{\Lambda}$.
The conditional distribution of $\alpha^2$ is 
\begin{equation} \label{eq:cond_alpha2}
P(\alpha^{2} \vert \boldsymbol{\Sigma}) \propto \alpha_{0}^{2} \text{tr}( \boldsymbol{\Lambda} \boldsymbol{\Sigma}^{-1}) / \chi^{2}_{\rho (J - 1)}.
\end{equation}

To complete the specification of the MNP model, we specify the prior $\boldsymbol{\beta} \sim N(\boldsymbol{0}, \boldsymbol{B}_{0}^{-1}$).
It is known that predictions under the Bayesian formulation of the MNP model can be sensitive to the selection of the base alternative $J$ \citep{burgette2012trace}. 

\subsection{Multinomial robit (MNR)}

\begin{figure}[H]
\centering
\includegraphics[width = \textwidth]{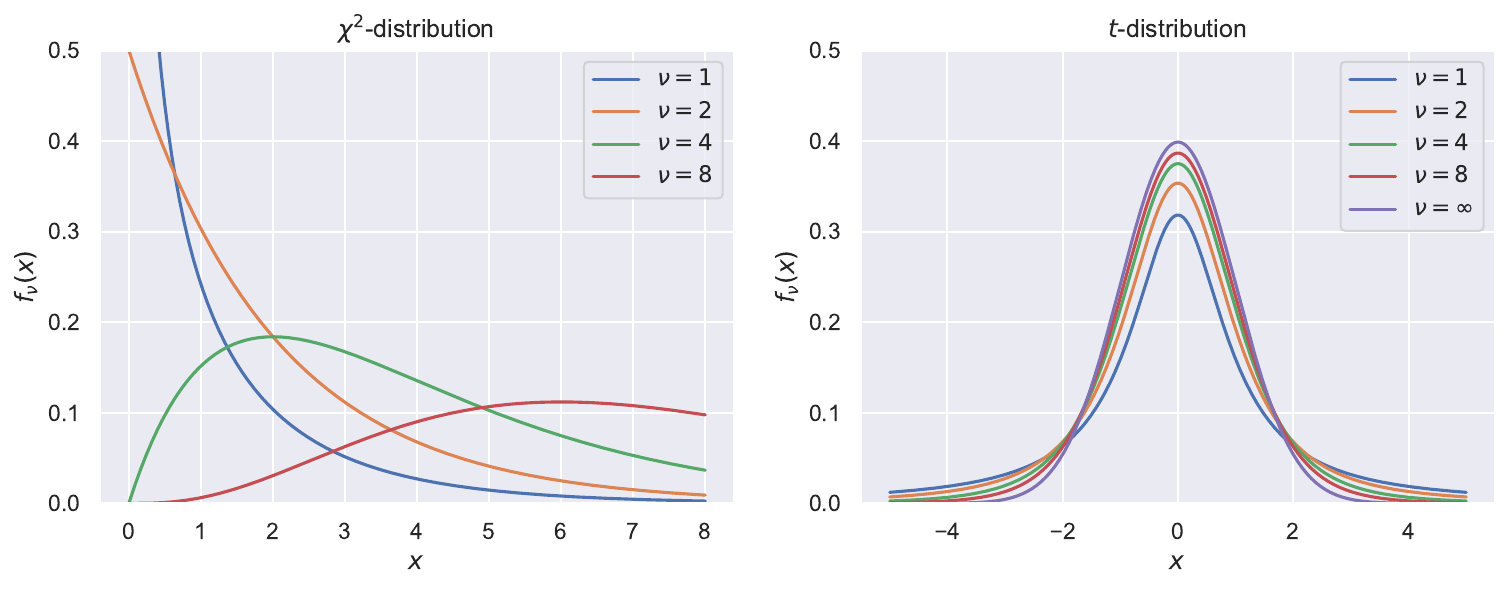}
\caption{Relationship between $\chi^2$- and t-distributions for different degrees of freedom $\nu$} \label{fig:chi2_t}
\end{figure}

The MNR model assumes a t-distributed kernel error for the latent variable $\boldsymbol{w}_{i}$, i.e. 
\begin{equation}
\boldsymbol{w}_{i} = \boldsymbol{X}_{i} \boldsymbol{\beta} + \boldsymbol{\varepsilon}_{i}
\quad
\text{with } \boldsymbol{\varepsilon}_{i} \sim t(\boldsymbol{0}, \boldsymbol{\Sigma}, \nu),
\quad \text{for } i = 1, \ldots, N,
\end{equation}
where $\boldsymbol{\Sigma}$ is a $(J - 1) \times (J - 1)$ covariance matrix and $\nu$ is a scalar DOF parameter.  
The t-distribution has the following normal mixture representation \citep{ding2014bayesian}:
\begin{equation} \label{eq:lv_mnr}
\boldsymbol{\varepsilon}_{i} \sim N(\boldsymbol{0}, \boldsymbol{\Sigma} / q_{i})
\quad 
\text{with } q_{i} \sim \chi^{2}_{\nu} / \nu, 
\quad \text{for } i = 1, \ldots, N.
\end{equation}
The latent variables $\boldsymbol{q} = \{ q_{1}, \ldots, q_{N} \}$ allow for heavy-tailedness in the distribution of the kernel error by increasing the variability of $\boldsymbol{\varepsilon}_{i}$ across different $i$. 
Figure \ref{fig:chi2_t} illustrates the relationship between the $\chi^2$-distribution (which controls the distribution of $\boldsymbol{q}$) and a $t$-distribution (which controls the distribution of $\boldsymbol{\varepsilon}$) with unit variance for different DOF $\nu$. For small $\nu < 30$, the t-distribution exhibits heavy tails. As $\nu$ approaches $\infty$, the $t$-distribution converges to the normal distribution.
We use the same priors for $\boldsymbol{\beta}$ and $\boldsymbol{\Sigma}$ as in MNP and also introduce the working parameter $\alpha$.
In addition, we put a Gamma prior on $\nu$ with $\nu \sim \text{Gamma}(\alpha_{0}, \beta_{0})$.
Predictions under the Bayesian formulation of the MNR model can be sensitive to the selection of the base alternative in the same way as predictions under the Bayesian formulation of the MNP model.

\subsection{Generalised multinomial robit (Gen-MNR)}

We generalise MNR by allowing for different marginal heavy-tailedness in the distribution of the latent variable $\boldsymbol{w}_{i}$. The Gen-MNR model assumes that the kernel error of $\boldsymbol{w}_{i}$ is drawn from a non-elliptically contoured $t$-distribution \citep[NECT;][]{jiang2016robust}. We have
\begin{equation} \label{eq:lv_genmnr}
\boldsymbol{w}_{i} = \boldsymbol{X}_{i} \boldsymbol{\beta} + \boldsymbol{\varepsilon}_{i}
\quad
\text{with } \boldsymbol{\varepsilon}_{i} \sim \text{NECT}_{\boldsymbol{p}}(\boldsymbol{0}, \boldsymbol{\Sigma}, \boldsymbol{\nu}),
\quad \text{for } i = 1, \ldots, N,
\end{equation}
where $\boldsymbol{\Sigma}$ is a $(J - 1) \times (J - 1)$ covariance matrix and $\boldsymbol{\nu} = \{ \nu_{1}, \ldots, \nu_{M} \}$ is a $M$ vector of DOF with $1 < M \leq J - 1$. $\boldsymbol{p} = \{ p_{1}, \ldots, p_{M} \}$ is a $M$ vector giving the number of dimensions that are associated with each DOF $\nu_{m}$. We have $p_{m} \in \mathbb{N} \setminus \{ 0 \}$ and $\sum_{m=1}^{M} p_{m} = J - 1$.
The NECT distribution has the following normal mixture representation \citep{jiang2016robust}:
\begin{equation}
\boldsymbol{\varepsilon}_{i} = \boldsymbol{Q}_{i}^{-1/2} \boldsymbol{\Sigma}^{1/2} \boldsymbol{Z}_{i},
\quad
\boldsymbol{Z}_{i} \sim N(\boldsymbol{0}, \boldsymbol{I}_{J-1}),
\quad \text{for } i = 1, \ldots, N,
\end{equation}
where $\boldsymbol{Q}_{i} = \text{diag}( q_{i1} \boldsymbol{I}_{p_{1}}, \ldots, q_{iM} \boldsymbol{I}_{p_{M}} \}$ is a $(J - 1) \times (J - 1)$ block-diagonal matrix with $q_{im} \sim \chi^{2}_{\nu_{m}} / \nu_{m}$ for $m = 1, \ldots, M$. $\boldsymbol{I}_{l}$ is a $l \times l$ identity matrix.
Each marginal component of a NECT-distributed random variable follows a univariate t-distribution with the respective DOF, i.e. if $\boldsymbol{\varepsilon} \sim \text{NECT}_{\boldsymbol{p}}(\boldsymbol{0}, \boldsymbol{\Sigma}, \boldsymbol{\nu})$, then $\varepsilon_{j} \sim t(0, \Sigma_{jj}, \nu_{m(j)})$, where $m(j)$ maps dimension $j$ onto its associated DOF. 
In the rest of this work, we assume that $M = J-1$ without loss of generality. 
The Gen-MNR model uses the same prior distributions as the MNR model and also includes the working parameter $\alpha$.
We let $\nu_{j} \sim \text{Gamma}(\alpha_{0}, \beta_{0})$ for $j = 1, \ldots, J-1$.
Predictions of the Gen-MNR model can be sensitive to the selection of the base alternative in the same way as predictions of the MNP and MNR models.

\section{Inference and implementation details} \label{sec:infer}

For the estimation of the MNP, MNR and Gen-MNR models, we employ Markov chain Monte Carlo methods in the form of Gibbs sampling \citep{robert2013monte}. 
The sampling schemes for MNP, MNR and Gen-MNR are presented in Algorithms \ref{algo:mnp}, \ref{algo:mnr} and \ref{algo:genmnr}, respectively.
Algorithm \ref{algo:mnp} is based on the Gibbs samplers proposed by \citet{burgette2012trace} and \citet{imai2005bayesian}.
Algorithms \ref{algo:mnr} and \ref{algo:genmnr} are extensions of Algorithm \ref{algo:mnp}.
Whereas Algorithm \ref{algo:mnr} incorporates elements of the Gibbs sampler proposed by \citet{ding2014bayesian} for the robust Heckman selection model, Algorithm \ref{algo:genmnr}  incorporates elements of the Gibbs sampler proposed by \citet{jiang2016robust} for the multivariate robit model. 

Data augmentation \citep{tanner1987calculation} facilitates the construction of the samplers. 
The central idea of Bayesian data augmentation is to treat latent variables as unknown model parameters, which are imputed in additional sampling steps. Each of the samplers uses the data augmentation scheme developed by \citet{albert1993bayesian} and \citet{mcculloch1994exact} to impute the latent variable $\boldsymbol{w}$ (see Appendix \ref{app:sampling_w} for details). 
The samplers for the MNR and Gen-MNR models additionally incorporate the data augmentation schemes devised by \citet{ding2014bayesian}  and \citet{jiang2016robust}, respectively, to impute the latent variable $\boldsymbol{q}$. 
Data augmentation circumvents complex likelihood calculations in the estimation of the MNP, MNR and Gen-MNR models. This is because conditional on $\boldsymbol{w}$ and $\boldsymbol{q}$ (if applicable), the models reduce to standard Bayesian linear models.

Furthermore, following \citet{imai2005bayesian}, we leverage marginal data augmentation \citep[MDA;][]{van2001art, van2010marginal} to increase the rate of convergence of the MCMC algorithms. 
The central idea of MDA is to marginalise out the working parameter $\alpha$, which is not identified given the observed data $\boldsymbol{y}$, but is identified given the augmented data $\{\boldsymbol{y}, \boldsymbol{w} \}$, in some of the conditional updates of the model parameters in order to improve mixing of the Markov chains. 
\citet{imai2005bayesian} argue that because $\alpha^{2}$ is assigned a distribution with positive variance in (\ref{eq:cond_alpha2}), the marginal model $\int P(\boldsymbol{y}, \boldsymbol{w} \vert \boldsymbol{\theta}, \alpha) p(\alpha \vert \boldsymbol{\theta}) d \alpha$ is more diffuse and thus mixes faster than the conditional model $P(\boldsymbol{y}, \boldsymbol{w} \vert \boldsymbol{\theta}, \alpha)$.

The full conditional distribution of $\nu$ in the MNR model as well as the full conditional distributions of $\nu_{j}$ and $q_{ij}$ in the Gen-MNR model are nonstandard. To draw from these intricate distributions, we use Metropolised Independence samplers \citep{liu2008monte} with approximate Gamma proposals, as devised by \citet{ding2014bayesian}  and \citet{jiang2016robust} (see Appendices \ref{app:sampling_nu} and \ref{app:sampling_q} for details). 

Draws from the univariate truncated normal distribution are generated using a combination of inverse transform sampling and rejection sampling with a Rayleigh proposal \citep{botev2016simulation}. This hybrid approach is adopted to warrant reliable and efficient posterior simulations. On its own, the inverse transform sampling method breaks down due to numerical issues when the truncation interval is too far in the tail. Rejection sampling remedies this problem by allowing for reliable simulation in the tail. The Rayleigh proposal has been shown to be faster than other proposal distributions. For a detailed analysis of sampling methods for the truncated normal distribution, the reader is directed to \citet{botev2016simulation}.

We implement the MCMC algorithms for all models considered in this paper in Julia \citep{bezanson2017julia}.\footnote{The estimation code is available at \url{https://github.com/RicoKrueger/robit}.} 
In the subsequent applications, the Gibbs samplers are executed with a single chain consisting of 100,000 draws including a warm-up period of 50,000 draws. A thinning factor of 10 is applied to the post warm-up draws. Convergence is assessed with the help of the potential scale reduction factor \citep{gelman1992inference}.

Predictive choice distributions are obtained using simulation since the considered models lack closed-form choice probabilities. 
Choice probabilities in MNP are calculated using the GHK simulator \citep{hajivassiliou1996simulation, train2009discrete}; 
choice probabilities in MNR and Gen-MNR obtained using frequency simulators \citep{lerman1981use, geweke1994alternative}.
Frequency simulators are fast and easy to implement and provide unbiased approximations of choice probabilities \citep{geweke1994alternative}. 
However, in some applications, frequency simulators may have a positive probability that the simulated choice probability of an alternative is zero for any finite number of draws \citep{train2009discrete}. 
Zero simulated choice probabilities are especially likely when the underlying true choice probability is small \citep{train2009discrete}. 
Zero probabilities create issues in the calculation of the log-likelihood because the logarithm of zero is undefined \citep{train2009discrete}. 
In the applications presented in this paper, we only encountered the problem of zero simulated choice probabilities in the MNP model. 
Therefore, we adopt the GHK simulator for the calculation of predictive choice distributions in MNP. The recursive GHK simulator does not suffer from the issue of zero simulated choice probabilities and has been shown to provide accurate and reliable approximations of probit choice probabilities \citep{geweke1994alternative}.

\begin{algorithm}[h]
\caption{Gibbs sampler for the MNP model}
\label{algo:mnp}
\begin{algorithmic}
\State \textbf{Step 0:} Initialise parameters $\boldsymbol{w}$, $\boldsymbol{\beta}$, $\boldsymbol{\Sigma}$, $\alpha$.
\For{$t = 1, \ldots, T$}
\State \textbf{Step 1:} Update $\boldsymbol{w}$, $\alpha^{2}$.
\For{$i=1, \ldots, N$}
\For{$j=1, \ldots , J-1$}
\State Draw $w_{ij} \vert \boldsymbol{w}_{i,-j}, \cdot$ as explained in Appendix \ref{app:sampling_w}.
\EndFor
\EndFor
\State Draw $\alpha^{2} \vert \cdot  \sim \text{tr} (\boldsymbol{\Lambda} \boldsymbol{\Sigma}^{-1}) / \chi^{2}_{\rho (J - 1)}$.
\State Set $\tilde{\boldsymbol{w}} = \alpha \boldsymbol{w}$.
\State \textbf{Step 2:} Update $\tilde{\boldsymbol{\beta}}$, $\alpha^{2}$.
\State Set $\hat{\boldsymbol{B}} = \left ( \sum_{i = 1}^{N} \boldsymbol{X}_{i}^{\top} \boldsymbol{\Sigma}^{-1} \boldsymbol{X}_{i} + \boldsymbol{B}_{0} \right )^{-1}$.
\State Set $\hat{\boldsymbol{\beta}} = \hat{\boldsymbol{B}} \left ( \sum_{i = 1}^{N} \boldsymbol{X}_{i}^{\top} \boldsymbol{\Sigma}^{-1} \tilde{\boldsymbol{w}}_{i} \right ) $.
\State Draw $\alpha^{2} \vert \cdot \sim \left ( \sum_{i=1}^{N} (\tilde{\boldsymbol{w}}_{i}  - \boldsymbol{X}_{i} \hat{\boldsymbol{\beta}})^{\top} \boldsymbol{\Sigma}^{-1} (\tilde{\boldsymbol{w}}_{i}  - \boldsymbol{X}_{i} \hat{\boldsymbol{\beta}}) + \hat{\boldsymbol{\beta}}^{\top} \boldsymbol{B}_{0} \hat{\boldsymbol{\beta}} + \text{tr} (\boldsymbol{\Lambda} \boldsymbol{\Sigma}^{-1}) \right ) / \chi^{2}_{(N + \rho) (J - 1)}$.
\State Draw $\tilde{\boldsymbol{\beta}} \vert \cdot \sim N( \hat{\boldsymbol{\beta}}, \alpha^{2} \hat{\boldsymbol{B}} )$.
\State \textbf{Step 3:} Update $\tilde{\boldsymbol{\Sigma}}$.
\State Set $\tilde{\boldsymbol{z}}_{i} = \tilde{\boldsymbol{w}}_{i} - \boldsymbol{X}_{i} \tilde{\boldsymbol{\beta}}$.
\State Draw $\tilde{\boldsymbol{\Sigma}} \vert \cdot \sim IW \left ( \rho + N, \tilde{\boldsymbol{\Lambda}} + \sum_{i = 1}^{N} \tilde{\boldsymbol{z}}_{i} \tilde{\boldsymbol{z}}_{i}^{\top} \right )$.
\State Set $\alpha^{2} = \text{tr}( \tilde{\boldsymbol{\Sigma}} ) / (J - 1)$.
\State Set $\boldsymbol{\Sigma} = \tilde{\boldsymbol{\Sigma}} / \alpha^{2}$, $\boldsymbol{w} = \tilde{\boldsymbol{w}} / \alpha$, $\boldsymbol{\beta} = \tilde{\boldsymbol{\beta}} / \alpha$.
\EndFor
\State \Return $\boldsymbol{\beta}$, $\boldsymbol{\Sigma}$
\end{algorithmic}
\end{algorithm}

\begin{algorithm}[h]
\caption{Gibbs sampler for the MNR model}
\label{algo:mnr}
\begin{algorithmic}
\State \textbf{Step 0:} Initialise parameters $\boldsymbol{w}$, $\boldsymbol{q}$, $\boldsymbol{\beta}$, $\boldsymbol{\Sigma}$, $\nu$.
\For{$t = 1, \ldots, T$}
\State \textbf{Step 1:} Update $\boldsymbol{q}$. 
\For{$i=1, \ldots, N$}
\State Set $\boldsymbol{z}_{i} = \boldsymbol{w}_{i} - \boldsymbol{X}_{i} \boldsymbol{\beta}$. 
\State Draw $q_{i} \vert \cdot \sim \chi^{2}_{\nu + J - 1} / \left ( \boldsymbol{z}_{i}^{\top} \boldsymbol{\Sigma}^{-1} \boldsymbol{z}_{i}  \right )$.
\EndFor
\State \textbf{Step 2:} Update $\nu$.
\State Calculate $\alpha^{*}$, $\beta^{*}$ as explained in Appendix \ref{app:sampling_nu}.
\State Draw proposal $\nu' \sim \text{Gamma}(\alpha^{*}, \beta^{*} )$.
\State Accept the proposal with probability $\min \left \{ 1, \exp \left ( l(\nu') - h(\nu') - l(\nu) + h(\nu) \right )  \right \}$, where $l(\nu)$ and $h(\nu)$ are defined in (\ref{eq:log_post_nu}) and (\ref{eq:log_gamma_nu}), respectively.
\State \textbf{Step 3:} Update $\boldsymbol{w}$, $\alpha^{2}$ as described in Algorithm~\ref{algo:mnp}, Step~1.
\State \textbf{Step 4:} Update $\tilde{\boldsymbol{\beta}}$, $\alpha^{2}$.
\State Set $\hat{\boldsymbol{B}} = \left ( \sum_{i = 1}^{N} q_{i} \boldsymbol{X}_{i}^{\top} \boldsymbol{\Sigma}^{-1} \boldsymbol{X}_{i} + \boldsymbol{B}_{0} \right )^{-1}$.
\State Set $\hat{\boldsymbol{\beta}} = \hat{\boldsymbol{B}} \left ( \sum_{i = 1}^{N} q_{i} \boldsymbol{X}_{i}^{\top} \boldsymbol{\Sigma}^{-1} \tilde{\boldsymbol{w}}_{i} \right ) $.
\State Draw $\alpha^{2} \vert \cdot \sim \left ( \sum_{i=1}^{N} q_{i} (\tilde{\boldsymbol{w}}_{i}  - \boldsymbol{X}_{i} \hat{\boldsymbol{\beta}})^{\top} \boldsymbol{\Sigma}^{-1} (\tilde{\boldsymbol{w}}_{i}  - \boldsymbol{X}_{i} \hat{\boldsymbol{\beta}}) + \hat{\boldsymbol{\beta}}^{\top} \boldsymbol{B}_{0} \hat{\boldsymbol{\beta}} + \text{tr} (\boldsymbol{\Lambda} \boldsymbol{\Sigma}^{-1}) \right ) / \chi^{2}_{(N + \rho) (J - 1)}$.
\State Draw $\tilde{\boldsymbol{\beta}} \vert \cdot \sim N( \hat{\boldsymbol{\beta}}, \alpha^{2} \hat{\boldsymbol{B}} )$.
\State \textbf{Step 5:} Update $\tilde{\boldsymbol{\Sigma}}$.
\State Set $\tilde{\boldsymbol{z}}_{i} = \tilde{\boldsymbol{w}}_{i} - \boldsymbol{X}_{i} \tilde{\boldsymbol{\beta}}$.
\State Draw $\tilde{\boldsymbol{\Sigma}} \vert \cdot \sim IW \left ( \rho + N, \tilde{\boldsymbol{\Lambda}} + \sum_{i = 1}^{N} q_{i} \boldsymbol{z}_{i} \boldsymbol{z}_{i}^{\top} \right )$.
\State Set $\alpha^{2} = \text{tr}( \tilde{\boldsymbol{\Sigma}} ) / (J - 1)$.
\State Set $\boldsymbol{\Sigma} = \tilde{\boldsymbol{\Sigma}} / \alpha^{2}$, $\boldsymbol{w} = \tilde{\boldsymbol{w}} / \alpha$, $\boldsymbol{\beta} = \tilde{\boldsymbol{\beta}} / \alpha$.
\EndFor
\State \Return $\boldsymbol{\beta}$, $\boldsymbol{\Sigma}$, $\nu$
\end{algorithmic}
\end{algorithm}

\begin{algorithm}[h]
\caption{Gibbs sampler for the Gen-MNR model}
\label{algo:genmnr}
\begin{algorithmic}
\State \textbf{Step 0:} Initialise parameters $\boldsymbol{w}$, $\boldsymbol{q}$, $\boldsymbol{\beta}$, $\boldsymbol{\Sigma}$, $\alpha$, $\boldsymbol{\nu}$.
\For{$t = 1, \ldots, T$}
\State \textbf{Step 1:} Update $\boldsymbol{q}$. 
\For{$i=1, \ldots, N$}
\For{$j=1, \ldots , J-1$}
\State Calculate $\alpha^{*}$, $\beta^{*}$ as explained in Appendix \ref{app:sampling_q}.
\State Draw proposal $q'_{ij} \sim \text{Gamma}(\alpha^{*}, \beta^{*} )$.
\State Accept the proposal with probability $\min \left \{ 1, \exp \left ( f(q'_{ij}) - g(q'_{ij}) - f(q_{ij}) + g(q_{ij}) \right )  \right \}$, where $f(q_{ij})$ and $g(q_{ij})$ are defined in (\ref{eq:log_post_q}) and (\ref{eq:log_gamma_q}), respectively.
\EndFor
\EndFor
\State \textbf{Step 2:} Update $\boldsymbol{\nu}$.
\For{$j = 1, \ldots, J - 1$}
\State Update $\nu_{j}$ as shown for $\nu$ in Algorithm~\ref{algo:mnp}, Step~2.
\EndFor
\State \textbf{Step 3:} Update $\boldsymbol{w}$, $\alpha^{2}$ as described in Algorithm~\ref{algo:mnp}, Step~1.
\State \textbf{Step 4:} Update $\tilde{\boldsymbol{\beta}}$, $\alpha^{2}$.
\State Set $\boldsymbol{Q}_{i} = \text{diag}(\boldsymbol{q}_{i})$.
\State Set $\hat{\boldsymbol{B}} = \left ( \sum_{i = 1}^{N} \boldsymbol{X}_{i}^{\top} \boldsymbol{Q}_{i}^{1/2} \boldsymbol{\Sigma}^{-1} \boldsymbol{Q}_{i}^{1/2} \boldsymbol{X}_{i} + \boldsymbol{B}_{0} \right )^{-1}$.
\State Set $\hat{\boldsymbol{\beta}} = \hat{\boldsymbol{B}} \left ( \sum_{i = 1}^{N} \boldsymbol{X}_{i}^{\top} \boldsymbol{Q}_{i}^{1/2} \boldsymbol{\Sigma}^{-1} \boldsymbol{Q}_{i}^{1/2} \tilde{\boldsymbol{w}}_{i} \right ) $.
\State Draw $\alpha^{2} \vert \cdot \sim \left ( \sum_{i=1}^{N} (\tilde{\boldsymbol{w}}_{i}  - \boldsymbol{X}_{i} \hat{\boldsymbol{\beta}})^{\top} \boldsymbol{Q}_{i}^{1/2} \boldsymbol{\Sigma}^{-1} \boldsymbol{Q}_{i}^{1/2} (\tilde{\boldsymbol{w}}_{i}  - \boldsymbol{X}_{i} \hat{\boldsymbol{\beta}}) + \hat{\boldsymbol{\beta}}^{\top} \boldsymbol{B}_{0} \hat{\boldsymbol{\beta}} + \text{tr} (\boldsymbol{\Lambda} \boldsymbol{\Sigma}^{-1}) \right ) / \chi^{2}_{(N + \rho) (J - 1)}$.
\State Draw $\tilde{\boldsymbol{\beta}} \vert \cdot \sim N( \hat{\boldsymbol{\beta}}, \alpha^{2} \hat{\boldsymbol{B}} )$.
\State \textbf{Step 5:} Update $\tilde{\boldsymbol{\Sigma}}$.
\State Set $\tilde{\boldsymbol{z}}_{i} = \tilde{\boldsymbol{w}}_{i} - \boldsymbol{X}_{i} \tilde{\boldsymbol{\beta}}$.
\State Draw $\tilde{\boldsymbol{\Sigma}} \vert \cdot \sim IW \left ( \rho + N,  \tilde{\boldsymbol{\Lambda}} + \sum_{i = 1}^{N} \boldsymbol{Q}_{i}^{1/2} \boldsymbol{z}_{i} \boldsymbol{z}_{i}^{\top} \boldsymbol{Q}_{i}^{1/2} \right )$.
\State Set $\alpha^{2} = \text{tr}( \tilde{\boldsymbol{\Sigma}} ) / (J - 1)$.
\State Set $\boldsymbol{\Sigma} = \tilde{\boldsymbol{\Sigma}} / \alpha^{2}$, $\boldsymbol{w} = \tilde{\boldsymbol{w}} / \alpha$, $\boldsymbol{\beta} = \tilde{\boldsymbol{\beta}} / \alpha$.
\EndFor
\State \Return $\boldsymbol{\beta}$, $\boldsymbol{\Sigma}$, $\boldsymbol{\nu}$
\end{algorithmic}
\end{algorithm}

\section{Simulation study} \label{sec:sim}

We consider two simulated examples to
i) assess the ability of the proposed Gibbs samplers to recover parameters in finite samples and
ii) illustrate the effect of ignoring the non-normality of the kernel error distribution on the sensitivities of the choice distributions to changes in an exogenous variable. 

\subsection{Example I: Data generated according to MNR model}

In the first example, we simulate data according to the generative process underlying the MNR model.
We let $N = 10,000$ and $J = 4$. We set 
$\boldsymbol{\beta} = ( -0.5, 0.5, -0.5, 1, -1 )^{\top}$, where the first three entries are bias terms pertaining to alternative-specific constants and the last two entries are utility parameters pertaining to alternative-specific attributes.
Furthermore, we let
$\boldsymbol{\Sigma} = 
\begin{bmatrix} 
1.0 & 0.3 & 0.0 \\
0.3 & 1.0 & 0.3 \\
0.0 & 0.3 & 1.0 \\
\end{bmatrix}$
and $\nu = 2$. 
Since the first three predictors are alternative-specific constants, we have $X_{ijk}^{\text{obs}} = 1$ for $i = 1, \ldots, N$, $j = 1, 2, 3$ and $k = 1, 2, 3$.
The remaining predictors are alternative-specific attributes. We draw $X_{ijk}^{\text{obs}} \sim U(0,4)$ for $i = 1, \ldots, N$, $j = 1, \ldots, J$ and $k = 4, 5$.
The fourth alternative is set as reference alternative both in data generation and model estimation. 
For the sake of simplicity, we do not perform a search over the specification of the reference alternative.
We employ the synthetic data to estimate MNP, MNR and Gen-MNR using the methods described in Section~\ref{sec:infer}.

\begin{figure}[H]
\centering
\includegraphics[width = 0.5 \textwidth]{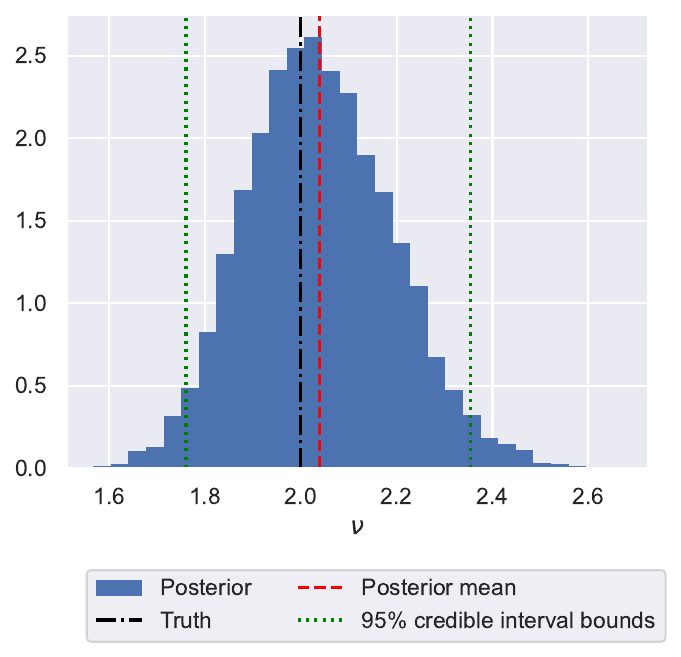}
\caption{Estimated posterior distribution and true value of the degree of freedom parameter $\nu$ for the MNR model in simulation example I} \label{fig:robit_sim_nu}
\end{figure}

Figure \ref{fig:robit_sim_nu} shows the marginal posterior distribution of the DOF parameter $\nu$ of the MNR model along with the corresponding true parameter value used in the generation of the data. 
It can be seen that Algorithm \ref{algo:mnr} performs well at recovering the DOF parameter of the MNR model since the true parameter value is well contained within the 95\% central credible interval. 
From Figures \ref{fig:robit_sim_beta} and \ref{fig:robit_sim_sigma} in Appendix \ref{app:sim1}, we can further conclude that Algorithm \ref{algo:mnr} also does an excellent job recovering the remaining parameters $\boldsymbol{\beta}$ and $\boldsymbol{\Sigma}$.

\begin{figure}[H]
\centering
\includegraphics[width = \textwidth]{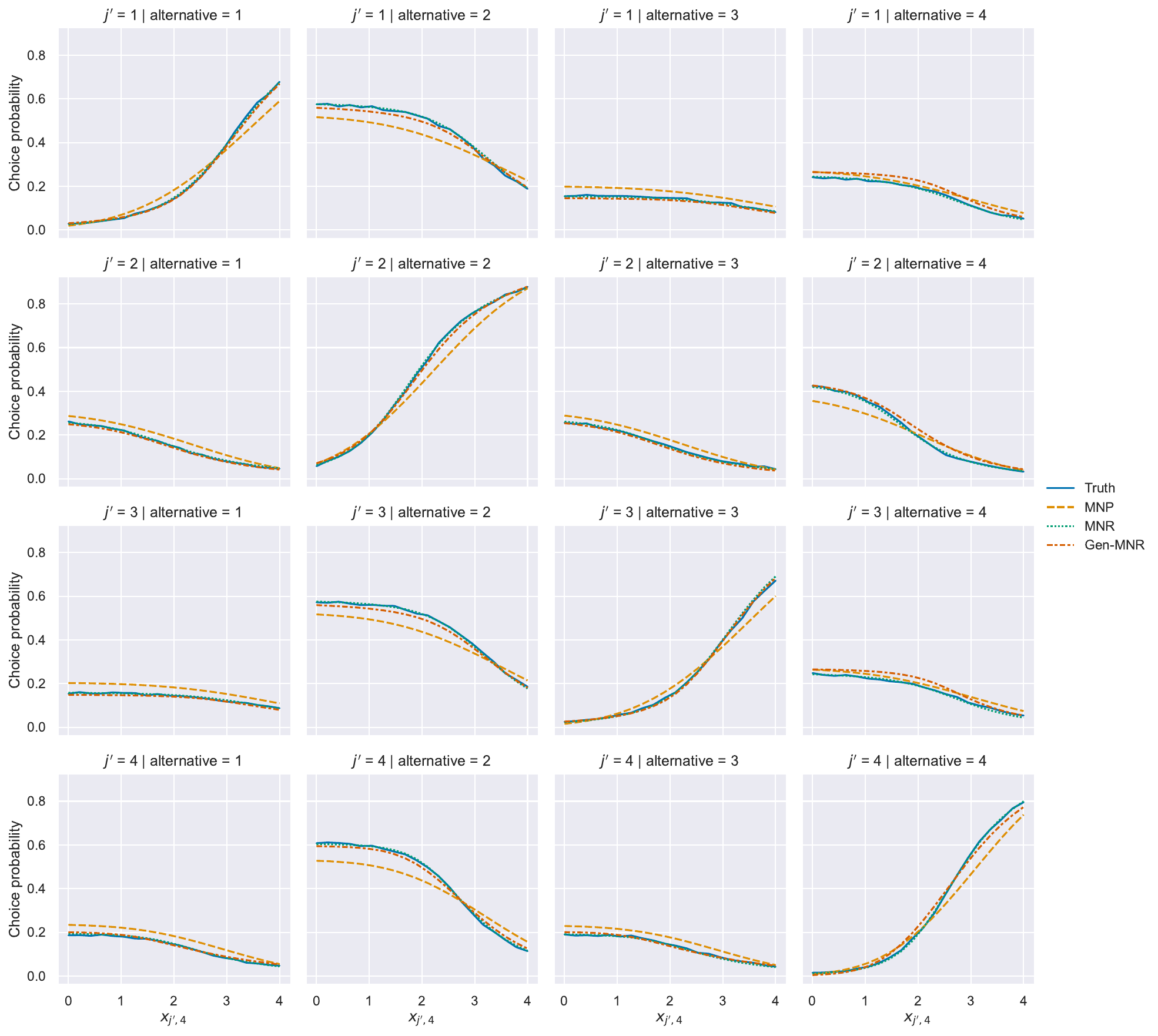}
\caption{Choice distribution sensitivities in simulation example I} \label{fig:sim1}
\end{figure}

Figure~\ref{fig:sim1} displays the choice distribution sensitivities of the estimated models and a ``true'' model, i.e.\ a MNR model with the parameter values used in the generation of the synthetic data.
To create Figure~\ref{fig:sim1}, we vary the first alternative-specific attribute of a selected alternative $j'$ over its range and predict the choice probabilities of all alternatives, while all other attributes are held fixed at their mean values.
Each row in Figure~\ref{fig:sim1} shows the posterior means of the predicted choice distributions as a function of attribute $X_{jk}^{\text{obs}}$ with $j = j'$ and $k=4$.
To summarise the results shown in Figure~\ref{fig:sim1} and to quantify the posterior uncertainty in the predictions, we also compute the posterior distributions of the Euclidean distances between the true and the predictive choice distribution sensitivities. We summarise these posterior distributions with mean, standard deviation and 95\%-credible intervals in Table~\ref{tab:sim1}.

From Figure~\ref{fig:sim1}, it can be seen that MNR and Gen-MNR are able to recover the choice distributions of the true model in all considered scenarios. 
By contrast, the MNP model, which is unable to capture the heavy-tailedness of the kernel error distribution, produces biased choice distributions.

These insights are corroborated by Table~\ref{tab:sim1}. 
In all considered scenarios, the posterior means of the Euclidean distances between the true and predicted choice distribution sensitivities are substantially smaller for MNR and Gen-MNR than for MNP. 
In nearly all considered scenarios, the credible intervals of the posterior distribution of the Euclidean distance between the true and predicted choice distribution sensitivity for MNP does not overlap with the credible intervals for MNR and Gen-MNR. 
In addition, the credible intervals of MNR and Gen-MNR overlap in all scenarios, which indicates that the differences between MNR and Gen-MNR are not statistically significant at a 5\%-significance level.

For example, for $j' = 4$ and alternative 2, we observe substantial differences between MNP and the true model Figure~\ref{fig:sim1}, while MNR and Gen-MNR appear to coincide with the true model. 
According to Table~\ref{tab:sim1}, the posterior mean of the Euclidean distance between the true model and MNP is 0.310, while the posterior mean of the Euclidean distance between the true model and MNR (Gen-MNR) is 0.048 (0.080). 
Table~\ref{tab:sim1} further suggests that for the same scenario, the credible interval of the posterior distribution of the Euclidean distance between the true and predicted choice distribution sensitivity for MNP does not overlap with the credible intervals for MNR and Gen-MNR, while the credible intervals for MNR and Gen-MNR coincide with each other.

In sum, these observations suggest that MNR and Gen-MNR perform significantly better than MNP at recovering the true choice distribution sensitivities.

\begin{table}[H]
\centering
\scriptsize
\input{table_sim1}
\caption{Posterior distributions of Euclidean distances between true and predicted choice distribution sensitivities in simulation example I} \label{tab:sim1}
\end{table}

\subsection{Example II: Data generated according to Gen-MNR model}

In the second example, we consider data generated according to the Gen-MNR model.
The data generating process is essentially same as in Example I, with the only difference we allow for different marginal heavy-tailedness by setting 
$\boldsymbol{\nu} = (5, 3, 1)^{\top}$.

\begin{figure}[H]
\centering
\includegraphics[width = \textwidth]{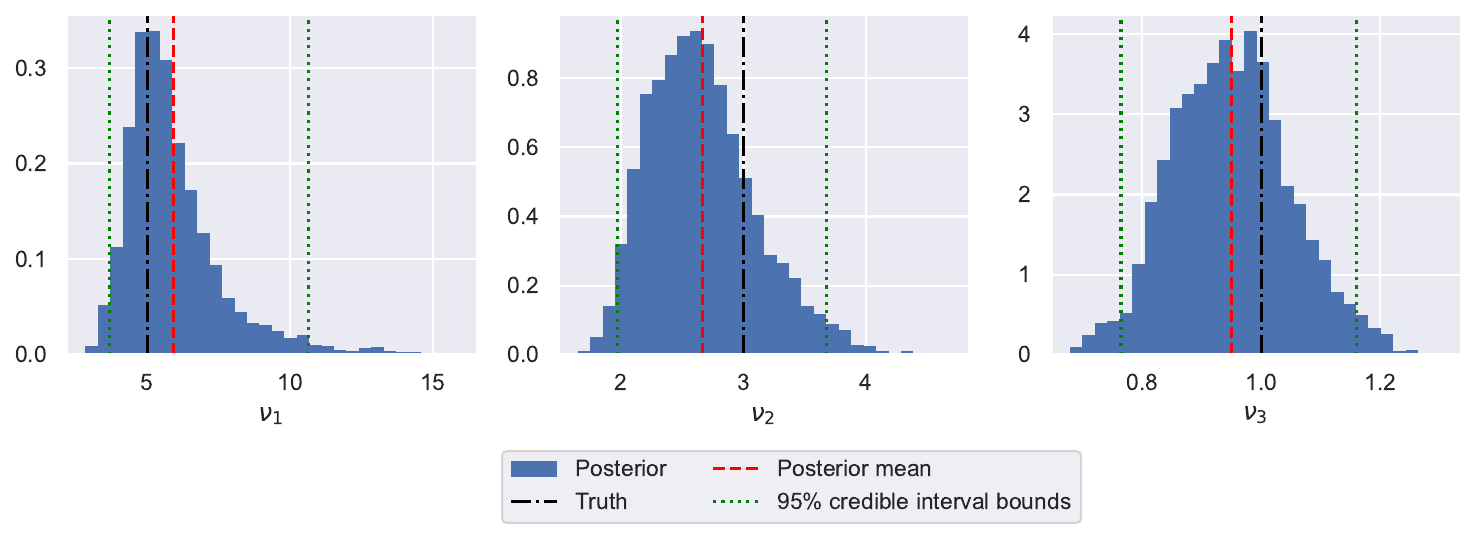}
\caption{Estimated posterior distributions and true values of the degree of freedom parameters $\nu_{1}$, $\nu_{2}$ and $\nu_{3}$ for Gen-MNR in simulation example II} \label{fig:genrobit_sim_nu}
\end{figure}

Figure \ref{fig:genrobit_sim_nu} shows the marginal posterior distributions of the DOF parameters $\nu_{1}$, $\nu_{2}$ and $\nu_{3}$ along with their corresponding true parameter values. 
It is evident that Algorithm \ref{algo:genmnr} is able to recover the DOF parameters of Gen-MNR since the true parameter values are all well contained within the 95\% central credible intervals.
Furthermore, Figures \ref{fig:genrobit_sim_beta} and \ref{fig:genrobit_sim_sigma} in Appendix \ref{app:sim2} show that Algorithm \ref{algo:genmnr} also performs well at recovering $\boldsymbol{\beta}$ and $\boldsymbol{\Sigma}$.

\begin{figure}[H]
\centering
\includegraphics[width = \textwidth]{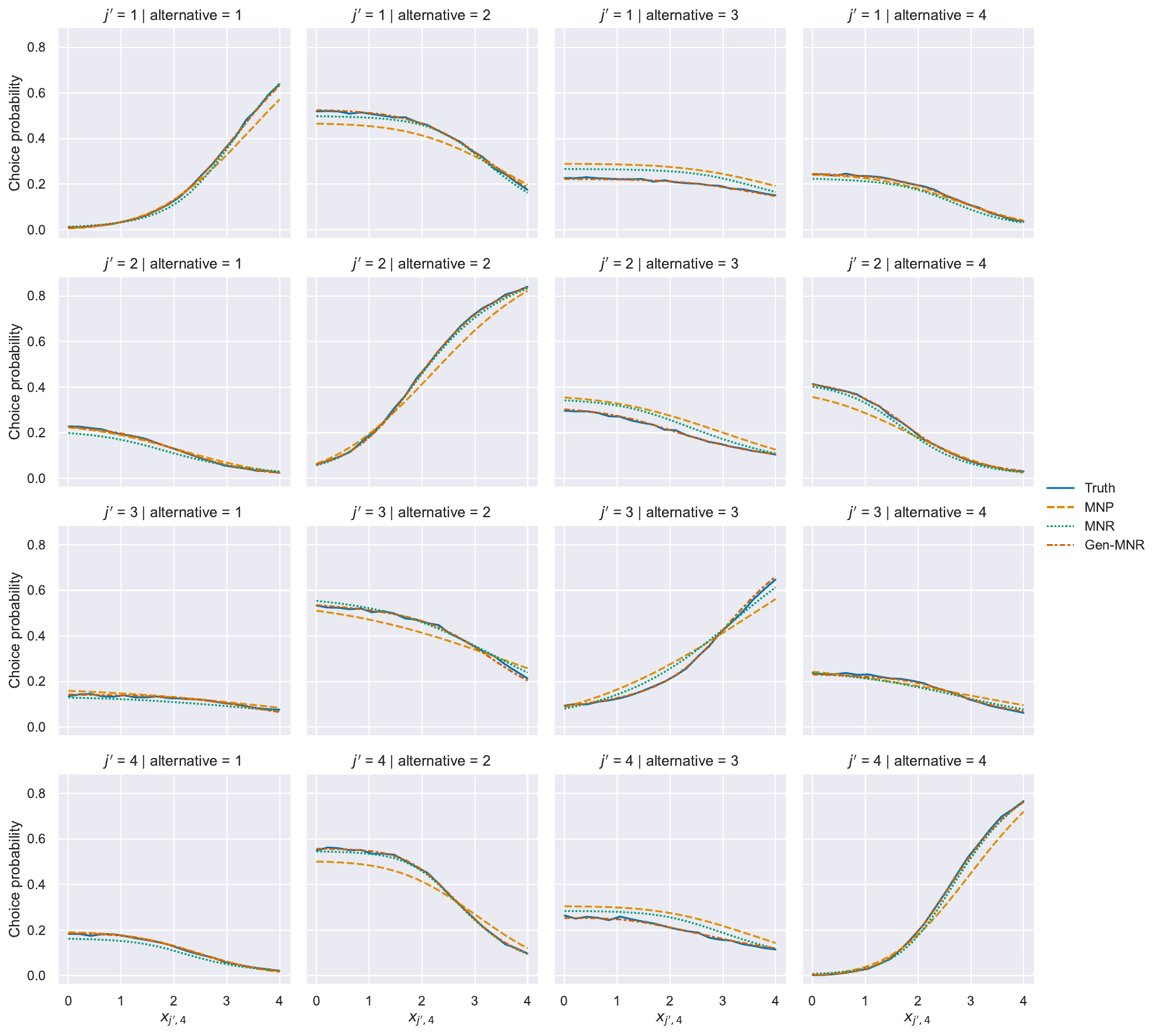}
\caption{Choice distribution sensitivity in simulation example II} \label{fig:sim2}
\end{figure}

Figure~\ref{fig:sim2} gives the choice distribution sensitivities for the estimated models and a ``true'' model, i.e.\ a Gen-MNR model with the parameter values used in the generation of the synthetic data.
Figure~\ref{fig:sim2} is created in the same way as Figure~\ref{fig:sim1} for Example I. 
Like in Example I, we also compute the posterior distributions of the Euclidean distances between the true and the predictive choice distribution sensitivities and summarise these posterior distributions with mean, standard deviation and 95\%-credible intervals (see Table~\ref{tab:sim2}).

Overall, Figure~\ref{fig:sim2} indicates that in Example II with distinct marginal heavy-tailedness, only Gen-MNR is able to recover the choice distributions of the true model in all scenarios. 
This is because MNR allows for distinct marginal heavy-tailedness, whereas MNR only allows for generic heavy-tailedness and MNP does not capture any heavy-tailedness. 

Upon closer inspection, Figure~\ref{fig:sim2} and Table~\ref{tab:sim2} highlight interesting differences in the abilities of the three models to recover choice probabilities.
The differences between the three models and the true model are relatively minor for alternative~1 (see the first column of Figure~\ref{fig:sim2}).
Table~\ref{tab:sim2} largely confirms this observation. 
For alternative~1, the differences between the three models are comparatively minor, as the credible intervals of the posterior distributions of the Euclidean distances between the true and predicted choice distribution sensitivities of the three models overlap in all scenarios except when $j’ = 1$.

For alternative~2, MNP is biased, while MNR and Gen-MNR perform well at recovering the choice distribution of the true model (see the second column of Figure~\ref{fig:sim2}). For alternative~2, Table~\ref{tab:sim2} shows that the posterior means of the Euclidean distances between the true and predictive choice distribution sensitivities are considerably larger for MNP than for MNR and Gen-MNR; also, the credible intervals of the posterior distribution of the Euclidean distances for MNP do not overlap with the credible intervals for MNR and Gen-MNR.

For alternative~3, only Gen-MNR is able to recover the choice distribution of the true model, whereas MNP and MNR produce biased choice distributions (see the third column of Figure~\ref{fig:sim2}).
Table~\ref{tab:sim2} shows that for alternative~3, the posterior means of the Euclidean distances between the true and predictive choice distribution sensitivities are considerably smaller for Gen-MNR than for MNP and MNR.
In addition, the credible intervals of the posterior distributions of the Euclidean distances for Gen-MNR do not overlap with the credible intervals for MNP and MNR, with the exception of a negligible overlap between credible intervals of Gen-MNR and MNR for $j’ = 3$.

For alternative~4, Figure~\ref{fig:sim2} and Table~\ref{tab:sim2} suggest that the differences between the three models are comparatively minor when $j' \in \{1, 3 \}$.
However, when $j' \in \{2, 4\}$, only MNR and Gen-MNR perform well at recovering the true choice distributions, while MNP is biased.

In sum, these patterns of differences are reflective of the distinct marginal heavy-tailedness of the kernel error distribution induced by the different elements of the degrees of freedom parameter $\boldsymbol{\nu}  = (\nu_{1}, \nu_{2}, \nu_{3})^{\top} = (5, 3, 1)^{\top}$.
Only Gen-MNR is able to account for the extreme heavy-tailedness caused by $\nu_{3} = 1$ in the true model.

\begin{table}[H]
\centering
\scriptsize
\input{table_sim2}
\caption{Posterior distributions of Euclidean distances between true and predicted choice distribution sensitivities in simulation example II} \label{tab:sim2}
\end{table}

\section{Case study} \label{sec:case}


\subsection{Data and utility specification}  \label{sec:data}

Revealed preference data for the case study are sourced from the London Passenger Mode Choice (LPMC) dataset \citep{hillel2018recreating}. 
The LPMC dataset consists of trip records from the London Travel Demand Survey, which was conducted from 2012 to 2015. 
For each trip record, \citet{hillel2018recreating} imputed tailored choice sets including the attributes of the chosen and the non-chosen alternatives using an online directions application programming interface. 
As the LPMC dataset is a revealed preference mode choice dataset derived from a household travel survey, we conjecture that the observed response data may be contaminated due to misreporting and misclassification. In addition, it is possible that random utility maximisation does not accurately represent the decision rules governing some of the observed choices. 

In this case study, we restrict our analysis to home-based trips reported by individuals who are at least 12 years old. The original dataset contains 58,584 of such observations from 26,904 individuals. For our analysis, we consider 10,820 observations from 5,000 randomly selected individuals for model training. In addition, we randomly select one observation each from an additional set of 1,250 randomly selected individuals for out-of-sample validation. There are four mode choice alternatives, namely walking, cycling, transit and driving. In the training data, the observed market shares are 16.4\%, 3.0\%, 37.8\% and 42.8\%, respectively. In the test data, the observed market shares are 14.5\%, 3.9\%, 41.8\% and 39.9\%, respectively.


Each of the three models uses the same specification of the systematic utility, as shown in Table~\ref{tab:utility_spec}.
The variable ``traffic variability'' is a measure of the driving travel time uncertainty for the given origin-destination pair. It is defined as the difference between the travel times in a pessimistic traffic scenario and in an optimistic one divided by the travel time in a typical, best-guess traffic scenario \citep[see][]{hillel2018recreating}. 

The drive alternative is set as reference alternative in the estimation of all models. We performed a search over the specification of the reference alternative but found no substantive differences in parameter estimates, in-sample fit and out-of-sample predictive accuracy for different specifications of the reference alternative.

\begin{table}[H]
\centering
\footnotesize
\input{table_utility_spec}
\caption{Utility specifications considered in case study} \label{tab:utility_spec}
\end{table}

\subsection{Results}

\subsubsection{In-sample fit and out-of-sample predictive accuracy}

Table \ref{tab:fit} compares the in-sample fit and out-of-sample predictive accuracy of MNP, MNR and Gen-MNR in terms of the log-likelihood evaluated at the posterior means of the model parameters and the Akaike information criterion (AIC).
We find that both MNR and Gen-MNR provide substantially higher log-likelihood values on the training data than MNP. 
Whereas MNP gives a log-likelihood of $-8297.2$ on the training data, MNR and Gen-MNR produce much higher log-likelihood values of $-8101.3$ and $-8092.8$, respectively.
Remarkably, MNR outperforms MNP by almost 200 log-likelihood points with just one more parameter. 
We also observe that MNR and Gen-MNR provide a slightly better out-of-sample accuracy than MNP.
In practice, Gen-MNR should be selected over MNP and MNR given that Gen-MNR provides the best in-sample fit and out-of-sample predictive accuracy out of the three considered models. Regardless of this recommendation, we contrast the three models in terms of parameter and elasticity estimates in the following subsections in order to highlight practical differences between the three models.

\begin{table}[H]
\centering
\input{table_cs_fit}
\caption{In-sample fit and out-of-sample predictive ability} \label{tab:fit}
\end{table}

\subsubsection{Parameter estimates}

In Table~\ref{tab:cs_results}, we summarise the posterior distributions of the parameters of MNP, MNR and Gen-MNR. 
For each parameter, we report the posterior mean, posterior standard deviation and bounds of the 95\% credible interval. 

The posterior distributions of the DOF parameters provide information about the heavy-tailedness of the kernel error distributions of MNR and Gen-MNR.
The posterior mean of the generic DOF parameter $\nu$ in MNR is 2.132,
which indicates that the heavy-tailedness of the respective kernel error distribution is substantial. 
The posterior distributions of the DOF parameters $\nu_{\text{walk}-\text{drive}}$, $\nu_{\text{cycle}-\text{drive}}$, $\nu_{\text{transit}-\text{drive}}$ in Gen-MNR suggest that the heavy-tailedness of the respective kernel error distribution is also  substantial, while the extent of heavy-tailedness varies significantly across the dimensions of the kernel error distribution.
Heavy-tailedness is most pronounced for the utility differences involving walking and cycling; the posterior means of the DOF parameters $\nu_{\text{walk}-\text{drive}}$, $\nu_{\text{cycle}-\text{drive}}$ are 4.737 and 1.399, respectively.
By contrast, the tails of the distribution of the utility differences between transit and driving are only moderately heavy, as the posterior mean of the respective DOF parameter $\nu_{\text{transit}-\text{drive}}$ is comparatively large with a value of 17.647.
The credible intervals of the DOF parameters in Gen-MNR are narrow, and only the credible intervals of the posterior distributions of $\nu_{\text{cycle}-\text{drive}}$ and $\nu_{\text{transit}-\text{drive}}$ overlap to a marginal extent.

All three models suggest that the considered alternative-specific attributes (see Table~\ref{tab:utility_spec}) have a statistically significant and negative influence on mode choice propensities.
Since the scales of the estimated utility parameters $\boldsymbol{\beta}$ are not guaranteed to be identical in the three models, the estimated effects cannot be compared in absolute terms.
Therefore, we contrast the sensitivities to alternative-specific attributes in terms of their implied willingness to pay (WTP), which is scale free and allows for a money-metric representation sensitivities. 
In general, WTP for an attribute $k$ is defined as the ratio of the marginal utility with respect to attribute $k$ and the marginal utility with respect to price. 
If the utility is specified as linear-in-parameters with only main effects as in the current application, WTP for an attribute $k$ is the given by the ratio $\beta_{k} / 
\beta_{p}$ of the non-price coefficient $\beta_{k}$ pertaining to the attribute of interest $k$ and the price coefficient $\beta_{p}$. 

Table \ref{tab:cs_wtp} summarises the posterior distributions of the WTP indicators in the three models.
WTP for reductions in out-of-vehicle travel time appears to be slightly larger in MNR and Gen-MNR than in MNP. 
Whereas the respective posterior means are 37.2 and 36.8 GBP/h in MNR and Gen-MNR, respectively, the respective posterior mean is only 32.7 GBP/h in MNP.
The posterior distributions of the WTP indicators for each of the remaining alternative-specific attributes are very similar across the three models. 

The models also provide insights into the influence of individual- and context-specific attributes on mode choice propensities. 
The differences between the three models are minor. 
For example, all models suggest that female travellers are relatively less likely to cycle and relatively more likely to use transit. 
While MNP and Gen-MNR suggest that women are more likely to select the driving mode, MNR does not suggest any gender differences in the propensity to use the driving mode.
In all three models, old age reduces the propensities to use transit and the driving mode. 
Furthermore, travel during winter months is found to reduce the propensity to cycle in all models. 
As expected, higher levels of car ownership increase the propensity to select the driving mode.

\begin{landscape}
\begin{table}[H]
\centering
\footnotesize
\input{table_cs_results}
\caption{Estimated parameters of MNP, MNR and Gen-MNR} \label{tab:cs_results}
\end{table}
\end{landscape}

\begin{table}[H]
\centering
\footnotesize
\input{table_cs_wtp}
\caption{Posterior distributions of willingness to pay indicators} \label{tab:cs_wtp}
\end{table}

\subsubsection{Elasticity estimates}

In Table~\ref{tab:cs_elas}, we enumerate selected aggregate arc elasticities for several policy-relevant scenarios. 
In principle, an aggregate arc elasticity $\eta^{W(j)}_{X_{j'k}}$ measures the responsiveness of the quantity demanded of an alternative $j$ by a group of observations relative to a change in attribute $k$ of alternative $j'$.
An aggregate arc elasticity as the ratio of the percentage change in the quantity demanded and the percentage change of the attribute.
Thus, we have 
\begin{equation}
\eta^{W(j)}_{X_{j'k}} = \frac{W^{*}(j) - W(j)}{W(j)} 
\frac{X_{j'k}}{X^{*}_{j'k} - X_{j'k}},
\end{equation}
where $W(j)$ and $W^{*}(j)$ denote the predicted market shares of alternative $j$ in the group under consideration before and after the change, respectively. 
Similarly, $X_{j'k}$ and $X^{*}_{j'k}$ denote the average of attribute $k$ of alternative $j'$ in the group before and after the change, respectively. 
In the current application, we compute aggregate arc elasticities for the whole sample. However, in principle, it is also possible to calculate aggregate arc elasticities for different exogenously defined subgroups of the sample.

Table~\ref{tab:cs_elas} shows that the aggregate arc elasticity estimates can differ substantially across the three models.
Overall, the results illustrate that Gen-MNR is more flexible than MNP and MNR. 
For example, the elasticities of the demand for cycling with respect to changes in cycling travel times differ quite considerably across the three models. 
For a 10\% decrease in the out-of-vehicle time of cycling, the own-elasticities of cycling are  $-0.94$, $-0.76$ and $-1.03$ in MNP, MNR and Gen-MNR, respectively. 
Reducing cycling travel times is an important policy-relevant scenario. 
For instance, the construction of cycling superhighways \citep[e.g.][]{rayaprolu2020impact} and a stimulation of e-bike uptake \citep[e.g.][]{dill2012electric} could result in widespread decreases of cycling travel times. 
Given the differences in the estimated aggregate arc elasticities, Gen-MNR makes a stronger case for the effectiveness of these interventions than MNP and MNR. 

The three models also produce different elasticities for changes to walking out-of-vehicle travel times.
The demand for walking is estimated to be more elastic to changes in walking time in MNR and Gen-MNR than in MNP.
More specifically, the aggregate arc elasticity of walking demand for a 10\% reduction in walking time is $-1.78$ in both MNR and Gen-MNR but is $-1.57$ in MNP. 
Besides, the demand for cycling is estimated to be more elastic to changes in walking time in Gen-MNR than in MNP and MNR.
The aggregate arc elasticity of cycling demand for a 10\% reduction in walking time is 0.27 in Gen-MNR, while it is 0.02 and 0.04 in MNP and MNR.
One possible way to reduce walking travel times are technological innovations such as fast-moving walkways \citep[e.g.][]{scarinci2017network}. 
Policies to support such walkways are more compelling under MNR and Gen-MNR estimates. 

Furthermore, we observe that the cross-elasticities of the demand for cycling with respect to changes in driving traffic variability, transit in-vehicle travel time and transit out-of-vehicle travel time differ markedly across the three models. 
For example, the aggregate arc elasticity of the demand for cycling with respect to a change in transit out-of-vehicle travel time is 0.31 in MNP and Gen-MNR but is 0.12 in MNR. 
Interestingly, in each of these three scenarios, the elasticity estimates of MNR are quite different from the elasticity estimates of MNP and Gen-MNR.
These differences are reflective of the rigidity imposed by the generic DOF parameter in MNR. 

\begin{table}[H]
\centering
\footnotesize
\input{table_cs_elas}
\caption{Aggregate arc elasticities} \label{tab:cs_elas}
\end{table}

\section{Conclusion} \label{sec:conc}

In this paper, we analysed two robust alternatives to the multinomial probit (MNP) model. 
Both alternatives belong to the family of robit models whose kernel error distributions are heavy-tailed t-distributions that moderate the influence of outliers. 
The first model is the multinomial robit (MNR) model, in which a single, generic degrees of freedom parameter controls the heavy-tailedness of the kernel error distribution. 
The second model, the generalised multinomial robit (Gen-MNR) model, is based on the non-elliptically contoured multivariate t-distribution, which allows for distinct heavy-tailedness in each dimension of the kernel error distribution. 
For both models, we devised Gibbs samplers for posterior inference. 

Using simulated data, we demonstrated the ability of the proposed Gibbs samplers for MNR and Gen-MNR to recover parameters in finite samples. 
We also showed that MNR and Gen-MNR outperform MNP at recovering choice distributions when the response data contain outliers. 
In a case study on transport mode choice behaviour, we demonstrated that MNR and Gen-MNR outperform MNP by considerable margins in terms of in-sample fit and out-of-sample predictive ability. 
We also illustrated that Gen-MNR provides the most flexible elasticity estimates. 

On the whole, our analysis suggests that Gen-MNR is a useful addition to the choice modelling toolbox due to its robustness properties.
In general, Gen-MNR should be preferred over the previously studied MNR model in daily data analysis practice because of its more flexible kernel error distribution.
Gen-MNR is also a parsimonious model. It requires estimating only $J - 2$ additional parameters compared to MNR (whereby $J$ denotes the number of alternatives).
In practice, the non-elliptical contoured t-distribution used in the formulation of Gen-MNR can also be specified in a way such that one DOF parameter controls the heavy-tailedness of more than one marginal of the kernel error distribution. Analysts can exploit this feature of Gen-MNR to achieve more parsimonious model specifications, especially in applications with many choice alternatives.

Our analysis suggests several directions for future research. 
First, hierarchical model extensions such as random parameters, latent variables and shrinkage priors could be considered.
Furthermore, aspects of the systematic parts of the latent utility differences could be represented using Bayesian additive regression trees (BART), which automatically partition large predictor spaces to capture interaction effects and nonlinearities \citep{chipman2010bart, kindo2016multinomial}. 
As these extensions are rooted in the Bayesian inferential paradigm, they can be incorporated into MNR and Gen-MNR with relative ease.
A second direction for future research is to use Bayesian modelling to automate aspects of the specification of MNR and Gen-MNR. 
For example, \citet{burgette2020symmetric} propose a symmetric prior which obviates the specification of a base alternative for MNP kernel error covariances. 
Such symmetric priors could also be developed for MNR and Gen-MNR.
Similarly, more diffuse prior formulations for the covariance of kernel error distributions of MNP, MNR and Gen-MNR can be investigated along the lines of \citet{huang2013simple}. 
In addition, a prior could be designed to automatically infer a parsimonious mapping of degrees of freedom parameters onto utility differences in the Gen-MNR. 
Finally, a third direction for future research is to formulate new robust discrete choice models using the skew-t-distribution \citep{kim2008flexible,lee2014finite}. The skew-t-distribution is an asymmetric probability distribution which can assume various smooth density shapes with either light or heavy tails and a longer tail either to the left or right. The skew-t-distribution is tractable in practical applications and parsimonious in the number of parameters controlling skewness \citep{kim2008flexible,lee2014finite}. Hence, discrete choice models based on the skew-t-distribution would afford further improvements in robustness relative to Gen-MNR which relies on a symmetric kernel error distribution. Developing MCMC algorithms for posterior inferences in discrete choice models based on the skew-t-distribution is also a promising direction for future research since efficient data augmentation schemes could be exploited in similar ways as in the current paper.

\section*{Author contribution statement}

RK: conception and design, method development and implementation, data processing and analysis, manuscript writing and editing, supervision. \\
MB: conception and design, manuscript editing, supervision. \\
TG: conception and design, method development and implementation, manuscript writing and editing. \\
PB: conception and design, manuscript writing and editing.

\section*{Conflicts of interest}

The authors declare that they have no conflicts of interests.

\newpage
\bibliographystyle{apalike}
\bibliography{bibliography.bib}

\newpage

\begin{appendices}

\section{Gibbs sampling details}

\subsection{\texorpdfstring{Sampling $\boldsymbol{w}$}{Sampling w}} \label{app:sampling_w}

To update $\boldsymbol{w}$, we iteratively sample from univariate truncated normal distributions. We have 
\begin{equation}
w_{ij} \sim TN(\mu_{ij}, \tau_{ij}^{2}), \quad \text{for } i = 1, \ldots, N, j = 1, \ldots, J-1.
\end{equation}
For MNP,
$\mu_{ij} = \boldsymbol{X}_{ij}^{\top} \boldsymbol{\beta} + \boldsymbol{\Sigma}_{j,-j} \boldsymbol{\Sigma}_{-j,-j}^{-1} (w_{i,-j} - \boldsymbol{X}_{i,-j} \boldsymbol{\beta})$  and
$\tau_{ij}^{2} = \boldsymbol{\Sigma}_{jj} - \boldsymbol{\Sigma}_{j,-j} \boldsymbol{\Sigma}_{-j,-j}^{-1} \boldsymbol{\Sigma}_{-j,j}$.
For MNR, 
$\mu_{ij} = \boldsymbol{X}_{ij}^{\top} \boldsymbol{\beta} + \boldsymbol{\Sigma}_{j,-j} \boldsymbol{\Sigma}_{-j,-j}^{-1} (w_{i,-j} - \boldsymbol{X}_{i,-j} \boldsymbol{\beta})$  and
$\tau_{ij}^{2} = (\boldsymbol{\Sigma}_{jj} - \boldsymbol{\Sigma}_{j,-j} \boldsymbol{\Sigma}_{-j,-j}^{-1} \boldsymbol{\Sigma}_{-j,j}) / q_{i}$.
For Gen-MNR,
$\mu_{ij} = \boldsymbol{X}_{ij}^{\top} \boldsymbol{\beta} + \boldsymbol{Q}_{ijj}^{-1/2} \boldsymbol{\Sigma}_{j,-j} \boldsymbol{\Sigma}_{-j,-j}^{-1} \boldsymbol{Q}_{i,-j,-j}^{1/2}  (w_{i,-j} - \boldsymbol{X}_{i,-j} \boldsymbol{\beta})$  and
$\tau_{ij}^{2} = (\boldsymbol{\Sigma}_{jj} - \boldsymbol{\Sigma}_{j,-j} \boldsymbol{\Sigma}_{-j,-j}^{-1} \boldsymbol{\Sigma}_{-j,j}) / q_{ij}$. 
Here, the index $-l$ denotes the vector without the $l$th element.
For all models, the constraint on $w_{ij}$ is 
$w_{ij} \geq \max \{ 0, w_{i,-j} \}$, if $y_{ij} = j$;
$w_{ij} < 0$, if $y_{ij} = J$;
$w_{ij} \leq \max \{ 0, w_{ij'} \}$, if $y_{ij} = j' \neq j$.

\subsection{\texorpdfstring{Sampling $\nu$}{Sampling nu}} \label{app:sampling_nu}

The full conditional distribution of $\nu$ is nonstandard. \citet{ding2014bayesian} shows that
\begin{equation} \label{eq:post_nu}
p(\nu \vert \cdot) 
\propto 
\exp \left \{
\frac{N \nu}{2} \log \left ( \frac{\nu}{2} \right) - N \log \Gamma \left ( \frac{\nu}{2} \right) + (\alpha_{0} - 1) \log \nu - \xi \nu
\right \},
\end{equation}
where
$\xi = \beta_{0} + \frac{1}{2} \sum_{i = 1}^{N} q_{i} - \frac{1}{2} \sum_{i = 1}^{N} \log q_{i}$. $\Gamma(x)$ denotes the Gamma function.
\citet{ding2014bayesian} proposes to sample from (\ref{eq:post_nu}) using a Metropolised Independence sampler \citep{liu2008monte} with an approximate Gamma proposal. 
The shape parameter $\alpha^{*}$ and the rate parameter $\beta^{*}$ of the proposal density are obtained as follows.  
The log conditional density of $\nu$ up to an additive constant is 
\begin{equation} \label{eq:log_post_nu}
l(\nu) = \frac{N \nu}{2} \log \left ( \frac{\nu}{2} \right) - N \log \Gamma \left ( \frac{\nu}{2} \right) + (\alpha_{0} - 1) \log \nu - \xi \nu.
\end{equation}
The log density of the Gamma proposal is
\begin{equation} \label{eq:log_gamma_nu}
h(\nu) = (\alpha^{*} - 1) \log \nu - \beta^{*} \nu.
\end{equation}
The first and second derivates of $l(\nu)$ and $h(\nu)$ are
\begin{equation} \label{eq:deriv1}
l'(\nu) = \frac{N}{2} \left [ \log \left ( \frac{\nu}{2} \right ) + 1 - \psi \left ( \frac{\nu}{2} \right ) \right ] + \frac{\alpha_{0} - 1}{\nu} - \xi,
\quad
h'(\nu) = \frac{\alpha^{*} - 1}{\nu} - \beta^{*},
\end{equation}
\begin{equation} \label{eq:deriv2}
l''(\nu) = \frac{N}{2} \left [ \frac{1}{\nu} -  \frac{1}{2} \psi' \left ( \frac{\nu}{2} \right ) \right ] + \frac{\alpha_{0} - 1}{\nu^{2}},
\quad
h''(\nu) = - \frac{\alpha^{*} - 1}{\nu^{2}},
\end{equation}
where $\psi(x)$ and $\psi'(x)$ are the di- and trigamma functions, respectively. 
The mode of $h(\nu)$ is $\frac{\alpha^{*} - 1}{\beta^{*}}$ and the corresponding curvature is $\frac{(\beta^{*})^{2}}{\alpha^{*} - 1}$.
We numerically find the mode $\nu^{*}$ of $l(\nu)$ and its corresponding curvature $l^{*} = l''(\nu^{*})$. 
Ultimately, we match the modes and the corresponding curvatures of $l(\nu)$ and $h(\nu)$ to obtain
\begin{equation}
\alpha^{*} = 1 - (\nu^{*})^{2} l^{*}, \quad \beta^{*} = - \nu^{*} l^{*}.
\end{equation}

\subsection{\texorpdfstring{Sampling $q_{ij}$}{Sampling q\textsubscript{ij}}} \label{app:sampling_q}

The full conditional distribution of $q_{ij}$ is nonstandard. \citet{jiang2016robust} show that
\begin{equation} \label{eq:post_q}
p(q_{ij} \vert \cdot) 
\propto 
\exp \left \{
- \frac{q_{ij} u_{ij}}{2} - \sqrt{q_{ij}} c_{ij} + \frac{\nu_{j} - 1}{2} \log q_{ij}
\right \},
\end{equation}
where 
$u_{ij} = \nu_{j} + ( \boldsymbol{\Sigma}^{-1} )_{jj} (w_{ij} - \boldsymbol{X}_{ij}^{\top} \boldsymbol{\beta})^{2}$ and
$c_{ij} = (w_{ij} - \boldsymbol{X}_{ij}^{\top} \boldsymbol{\beta}) \sum_{j' \neq j} \left ( \sqrt{ q_{ij'}  ( \boldsymbol{\Sigma}^{-1} )_{jj'}} (w_{ij} - \boldsymbol{X}_{ij}^{\top} \boldsymbol{\beta})  \right )$.
\citet{jiang2016robust} propose to sample from (\ref{eq:post_q}) using a Metropolised Independence sampler \citep{liu2008monte} with an approximate Gamma proposal. 
The shape parameter $\alpha^{*}$ and the rate parameter $\beta^{*}$ of the proposal density are obtained as follows.  
For $\nu_{j} \leq 1$, we set $\alpha^{*} = 1$ and $\beta^{*} = \frac{u_{ij}}{2}$.
For $\nu_{j} > 1$, $\alpha^{*}$ and $\beta^{*}$ are obtained through matching the modes and the corresponding curvatures of the target and the proposal densities. 
The log conditional density of $q_{ij}$ up to an additive constant is 
\begin{equation} \label{eq:log_post_q}
f(q_{ij}) = - \frac{q_{ij} u_{ij}}{2} - \sqrt{q_{ij}} c_{ij} + \frac{\nu_{j} - 1}{2} \log q_{ij}.
\end{equation}
The log density of the Gamma proposal is
\begin{equation} \label{eq:log_gamma_q}
g(q_{ij} ) = (\alpha^{*} - 1) \log q_{ij} - \beta^{*} q_{ij}.
\end{equation}
The mode of (\ref{eq:log_gamma_q}) and its corresponding curvature are $\frac{\alpha^{*} - 1}{\beta^{*}} = m_{ij}^{*}$ and $\frac{(\beta^{*})^{2}}{\alpha^{*} - 1} = l_{ij}^{*}$, respectively.
The first and second derivatives of (\ref{eq:log_post_q}) are
\begin{equation} 
f'(q_{ij}) = - \frac{u_{ij}}{2} - \frac{c_{ij}}{2 \sqrt{q_{ij}}} + \frac{\nu_{j} - 1}{2 q_{ij}},
\quad
f''(q_{ij}) = \frac{c_{ij}}{4 \sqrt{q_{ij}^{3}}} - \frac{\nu_{j} - 1}{2 q_{ij}^{2}}.
\end{equation}
The mode of (\ref{eq:log_post_q}) is $m_{ij}^{*} = \left( \frac{ \frac{c_{ij}}{2} + \sqrt{ \left (\frac{c_{ij}}{2} \right )^{2} + u_{ij} (\nu_{j} - 1)}}{\nu_{j} - 1} \right )^{-2}$,
and the corresponding curvature is $l_{ij}^{*} = f''(m_{ij}^{*})$.
After matching the modes and corresponding curvatures of the log target and the log proposal densities, we obtain
\begin{equation}
\alpha^{*} = 1 - (m_{ij}^{*})^{2} l_{ij}^{*}, \quad \beta^{*} = - m_{ij}^{*} l_{ij}^{*}.
\end{equation}

\section{Additional results for the simulation study}

\subsection{Example I} \label{app:sim1}

\begin{figure}[H]
\centering
\includegraphics[width = 0.6 \textwidth]{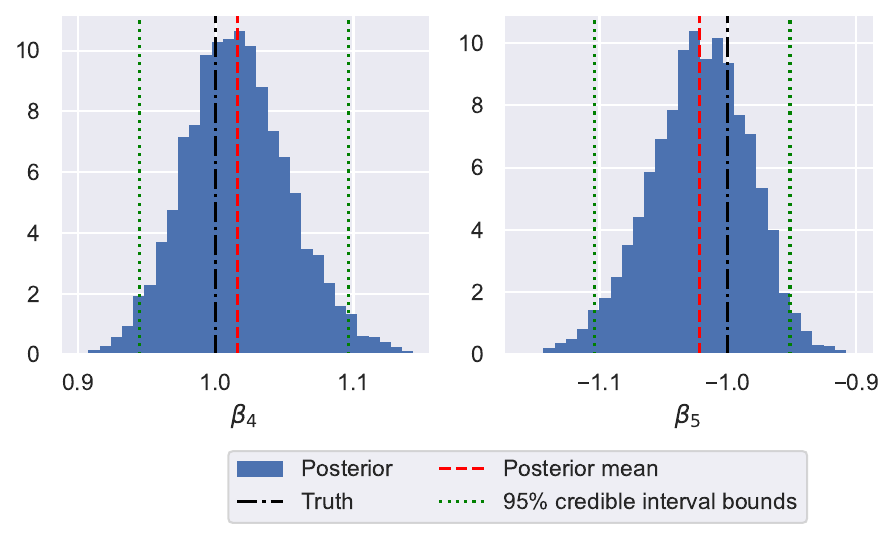}
\caption{Estimated posterior distribution and true values of the taste parameters $\{ \beta_{4}, \beta_{5} \}$ for MNR in simulation example I} \label{fig:robit_sim_beta}
\end{figure}

\begin{figure}[H]
\centering
\includegraphics[width = 0.8 \textwidth]{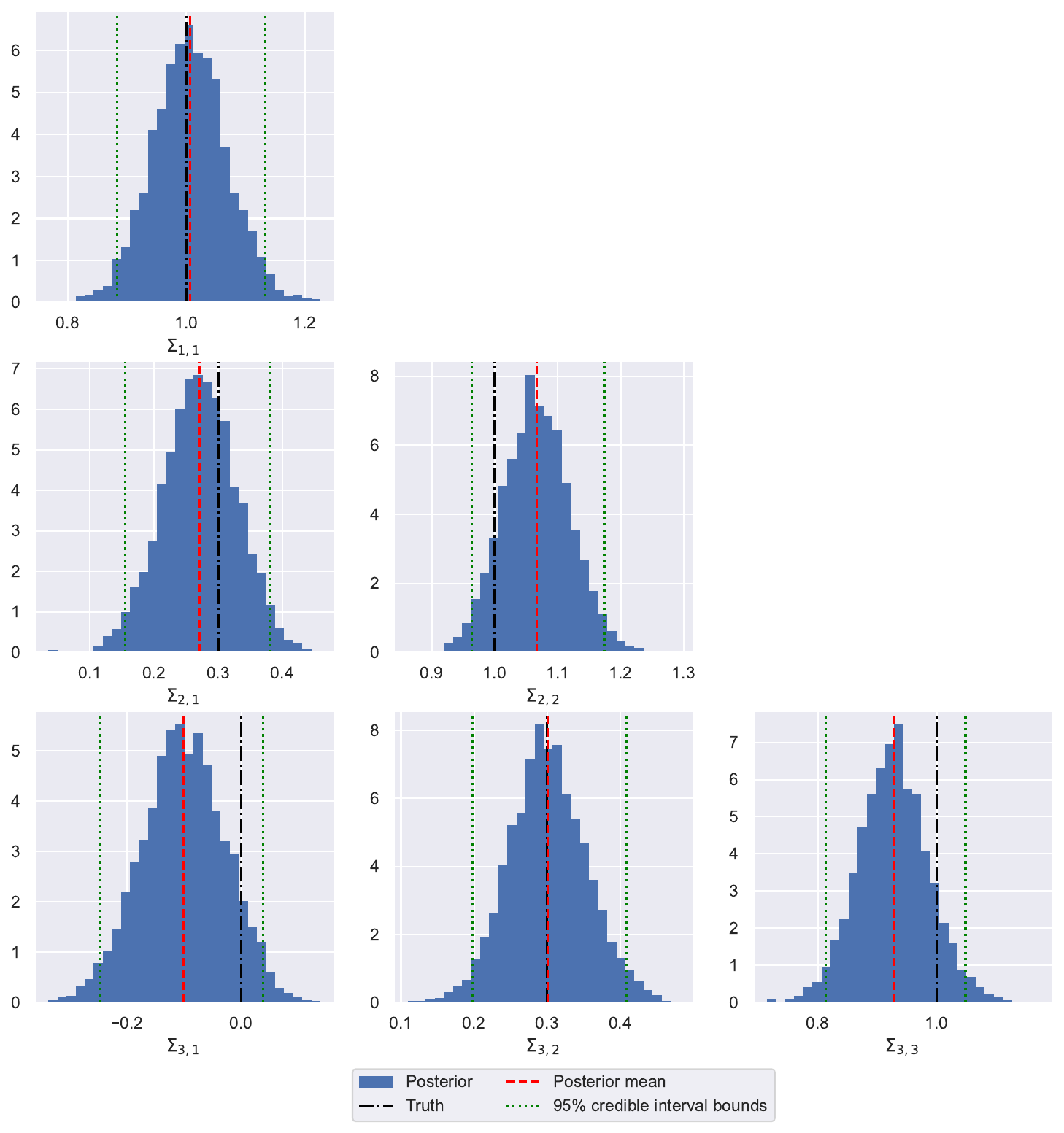}
\caption{Estimated posterior distribution and true values of the unique elements of the covariance matrix $\boldsymbol{\Sigma}$ for MNR in simulation example I} \label{fig:robit_sim_sigma}
\end{figure}

\subsection{Example II} \label{app:sim2}

\begin{figure}[H]
\centering
\includegraphics[width = 0.6 \textwidth]{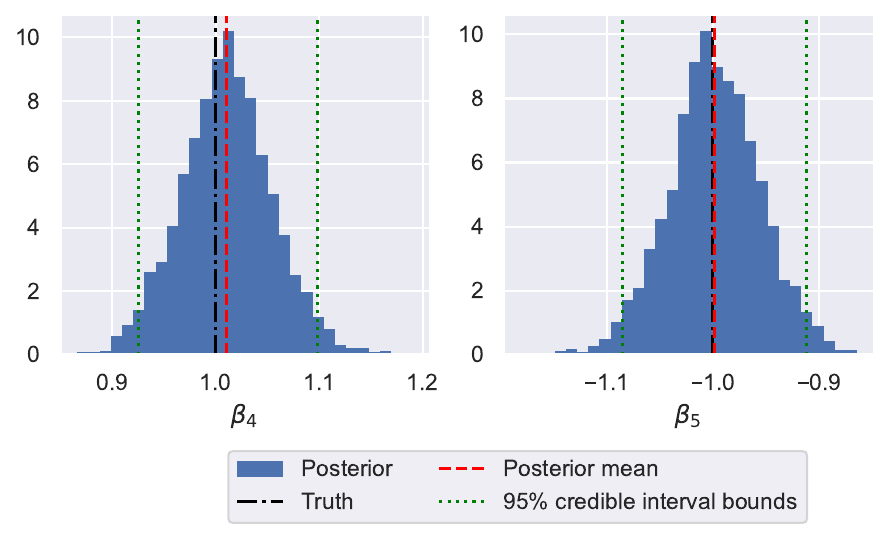}
\caption{Estimated posterior distribution and true values of the taste parameters $\{ \beta_{4}, \beta_{5} \}$ for the Gen-MNR model in simulation example II} \label{fig:genrobit_sim_beta}
\end{figure}

\begin{figure}[H]
\centering
\includegraphics[width = 0.8 \textwidth]{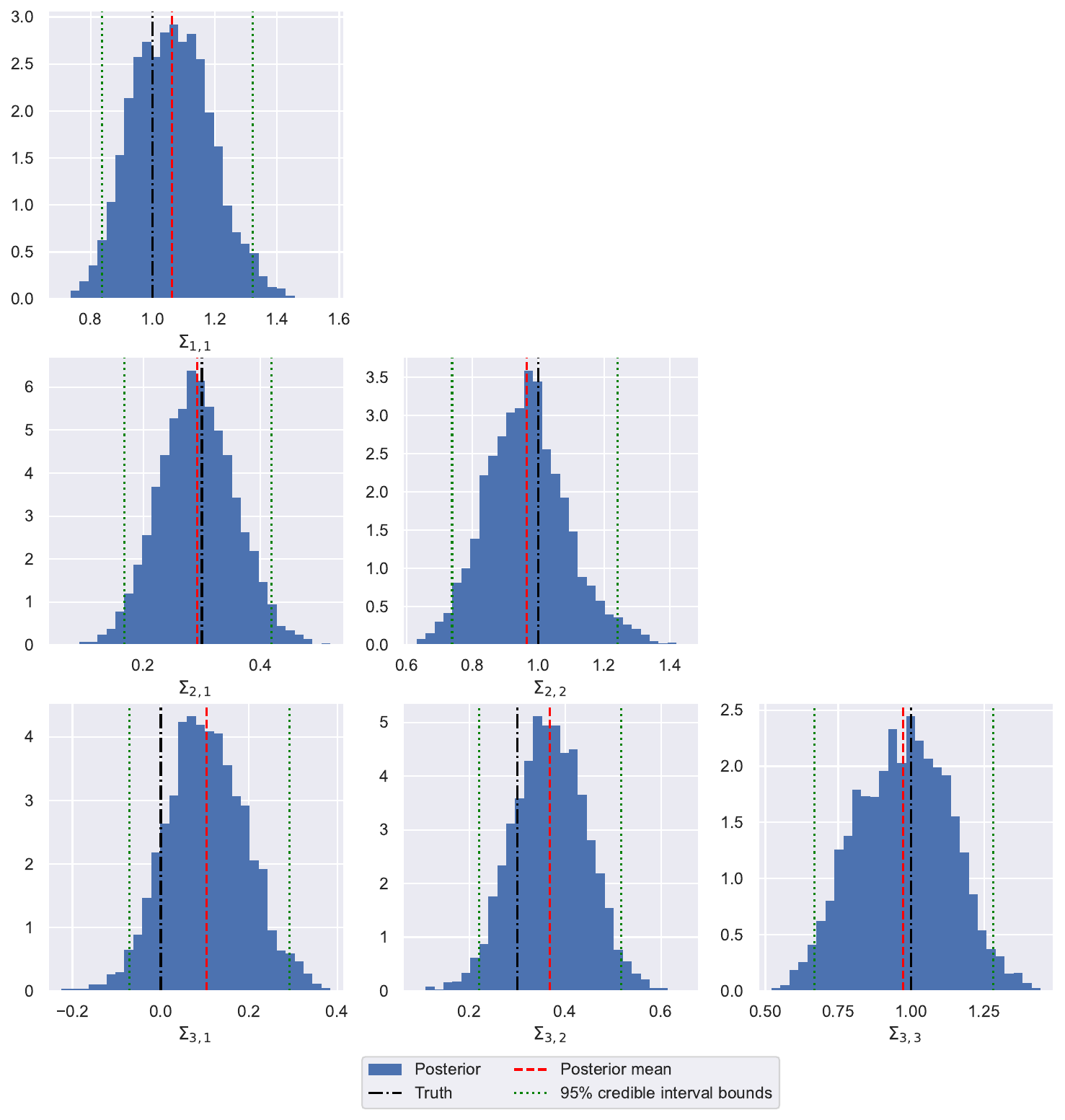}
\caption{Estimated posterior distribution and true values of the unique elements of the covariance matrix $\boldsymbol{\Sigma}$ for the Gen-MNR model in simulation example II} \label{fig:genrobit_sim_sigma}
\end{figure}

\end{appendices}

\end{document}

%% file: table_sim1.tex
\begin{tabular}{ll|rrrr|rrrr|rrrr}
\toprule
 &  & \multicolumn{4}{c|}{\textbf{MNP}} & \multicolumn{4}{c|}{\textbf{MNR}} & \multicolumn{4}{c}{\textbf{Gen-MNR}} \\
\textbf{$j'$} & \textbf{Alt.}  & \textbf{Mean} & \textbf{Std.} & \textbf{[2.5\%} & \textbf{97.5\%]} & \textbf{Mean} & \textbf{Std.} & \textbf{[2.5\%} & \textbf{97.5\%]} & \textbf{Mean} & \textbf{Std.} & \textbf{[2.5\%} & \textbf{97.5\%]} \\
\midrule
1 & 1 & 0.192 & 0.017 & 0.161 & 0.228 & 0.042 & 0.014 & 0.023 & 0.078 & 0.052 & 0.019 & 0.026 & 0.098 \\
 & 2 & 0.266 & 0.023 & 0.222 & 0.312 & 0.054 & 0.019 & 0.029 & 0.099 & 0.084 & 0.027 & 0.041 & 0.140 \\
 & 3 & 0.145 & 0.022 & 0.103 & 0.189 & 0.035 & 0.014 & 0.018 & 0.069 & 0.053 & 0.024 & 0.020 & 0.107 \\
 & 4 & 0.102 & 0.022 & 0.062 & 0.145 & 0.041 & 0.014 & 0.021 & 0.075 & 0.119 & 0.034 & 0.054 & 0.188 \\
\midrule
2 & 1 & 0.130 & 0.021 & 0.090 & 0.173 & 0.035 & 0.013 & 0.018 & 0.067 & 0.043 & 0.018 & 0.020 & 0.088 \\
 & 2 & 0.239 & 0.020 & 0.200 & 0.281 & 0.041 & 0.013 & 0.023 & 0.072 & 0.066 & 0.019 & 0.036 & 0.109 \\
 & 3 & 0.112 & 0.020 & 0.075 & 0.155 & 0.037 & 0.013 & 0.018 & 0.067 & 0.052 & 0.019 & 0.024 & 0.097 \\
 & 4 & 0.199 & 0.018 & 0.167 & 0.234 & 0.047 & 0.019 & 0.024 & 0.093 & 0.102 & 0.025 & 0.057 & 0.155 \\
\midrule
3 & 1 & 0.167 & 0.022 & 0.124 & 0.211 & 0.033 & 0.012 & 0.019 & 0.063 & 0.044 & 0.018 & 0.021 & 0.090 \\
 & 2 & 0.267 & 0.023 & 0.222 & 0.315 & 0.049 & 0.018 & 0.025 & 0.093 & 0.085 & 0.030 & 0.035 & 0.148 \\
 & 3 & 0.156 & 0.018 & 0.122 & 0.192 & 0.050 & 0.020 & 0.021 & 0.097 & 0.048 & 0.017 & 0.022 & 0.084 \\
 & 4 & 0.092 & 0.023 & 0.047 & 0.138 & 0.044 & 0.015 & 0.022 & 0.078 & 0.119 & 0.033 & 0.057 & 0.187 \\
\midrule
4 & 1 & 0.167 & 0.021 & 0.126 & 0.210 & 0.033 & 0.014 & 0.016 & 0.070 & 0.046 & 0.013 & 0.028 & 0.082 \\
 & 2 & 0.310 & 0.020 & 0.271 & 0.350 & 0.048 & 0.016 & 0.026 & 0.087 & 0.080 & 0.025 & 0.038 & 0.133 \\
 & 3 & 0.137 & 0.021 & 0.097 & 0.181 & 0.035 & 0.012 & 0.019 & 0.064 & 0.046 & 0.013 & 0.028 & 0.078 \\
 & 4 & 0.235 & 0.024 & 0.190 & 0.282 & 0.039 & 0.013 & 0.020 & 0.070 & 0.097 & 0.019 & 0.059 & 0.134 \\
\bottomrule
\end{tabular}

%% file: table_sim2.tex
\begin{tabular}{ll|rrrr|rrrr|rrrr}
\toprule
 &  & \multicolumn{4}{c|}{\textbf{MNP}} & \multicolumn{4}{c|}{\textbf{MNR}} & \multicolumn{4}{c}{\textbf{Gen-MNR}} \\
\textbf{$j'$} & \textbf{Alt.}  & \textbf{Mean} & \textbf{Std.} & \textbf{[2.5\%} & \textbf{97.5\%]} & \textbf{Mean} & \textbf{Std.} & \textbf{[2.5\%} & \textbf{97.5\%]} & \textbf{Mean} & \textbf{Std.} & \textbf{[2.5\%} & \textbf{97.5\%]} \\
\midrule
1 & 1 & 0.145 & 0.023 & 0.098 & 0.188 & 0.062 & 0.017 & 0.033 & 0.095 & 0.038 & 0.014 & 0.020 & 0.072 \\
 & 2 & 0.197 & 0.025 & 0.151 & 0.248 & 0.074 & 0.024 & 0.040 & 0.128 & 0.050 & 0.018 & 0.028 & 0.094 \\
 & 3 & 0.265 & 0.023 & 0.223 & 0.309 & 0.173 & 0.026 & 0.123 & 0.224 & 0.041 & 0.017 & 0.020 & 0.086 \\
 & 4 & 0.048 & 0.016 & 0.027 & 0.084 & 0.090 & 0.026 & 0.041 & 0.144 & 0.040 & 0.017 & 0.019 & 0.082 \\
\midrule
2 & 1 & 0.041 & 0.011 & 0.022 & 0.065 & 0.094 & 0.024 & 0.050 & 0.141 & 0.034 & 0.013 & 0.017 & 0.066 \\
 & 2 & 0.200 & 0.023 & 0.157 & 0.246 & 0.054 & 0.018 & 0.027 & 0.095 & 0.043 & 0.014 & 0.023 & 0.076 \\
 & 3 & 0.245 & 0.022 & 0.205 & 0.290 & 0.175 & 0.024 & 0.127 & 0.224 & 0.042 & 0.016 & 0.021 & 0.081 \\
 & 4 & 0.168 & 0.022 & 0.125 & 0.210 & 0.068 & 0.024 & 0.028 & 0.119 & 0.041 & 0.016 & 0.021 & 0.079 \\
\midrule
3 & 1 & 0.051 & 0.016 & 0.026 & 0.087 & 0.065 & 0.021 & 0.029 & 0.110 & 0.038 & 0.013 & 0.020 & 0.069 \\
 & 2 & 0.164 & 0.021 & 0.126 & 0.209 & 0.074 & 0.018 & 0.050 & 0.117 & 0.057 & 0.018 & 0.030 & 0.101 \\
 & 3 & 0.213 & 0.011 & 0.190 & 0.235 & 0.124 & 0.015 & 0.093 & 0.154 & 0.050 & 0.019 & 0.024 & 0.096 \\
 & 4 & 0.086 & 0.009 & 0.071 & 0.107 & 0.072 & 0.018 & 0.048 & 0.116 & 0.046 & 0.018 & 0.021 & 0.089 \\
\midrule
4 & 1 & 0.030 & 0.011 & 0.017 & 0.057 & 0.078 & 0.022 & 0.035 & 0.120 & 0.031 & 0.012 & 0.017 & 0.060 \\
 & 2 & 0.210 & 0.020 & 0.172 & 0.252 & 0.054 & 0.019 & 0.028 & 0.098 & 0.048 & 0.016 & 0.026 & 0.086 \\
 & 3 & 0.238 & 0.022 & 0.200 & 0.281 & 0.142 & 0.025 & 0.095 & 0.192 & 0.044 & 0.015 & 0.024 & 0.079 \\
 & 4 & 0.214 & 0.025 & 0.167 & 0.262 & 0.068 & 0.022 & 0.033 & 0.116 & 0.039 & 0.016 & 0.019 & 0.078 \\
\bottomrule
\end{tabular}

%% file: table_utility_spec.tex
\begin{tabular}{l|llll}
\toprule
\textbf{Variable} &  \textbf{Walk} &  \textbf{Cycle} &  \textbf{Transit} &   \textbf{Drive} \\
\midrule
Alternative-specific constants & & $\beta_{\text{asc, cycle}}$ & $\beta_{\text{asc, transit}}$ & $\beta_{\text{asc, drive}}$ \\
\midrule
Alternative-specific attributes \\
\quad Cost [GBP] & & & $\beta_{\text{cost}}$ & $\beta_{\text{cost}}$ \\
\quad Out-of-vehicle time (ovtt) [hours] & $\beta_{\text{ovtt}}$ & $\beta_{\text{ovtt}}$ & $\beta_{\text{ovtt}}$ \\
\quad In-vehicle travel time (ivtt) [hours] & & & $\beta_{\text{ivtt}}$ & $\beta_{\text{ivtt}}$ \\
\quad No. of transfers & & & $\beta_{\text{transfers}}$ \\
\quad Traffic variability (tv) & & & & $\beta_{\text{tv}}$ \\
\midrule
Individual- and context-specific attributes \\
\quad Female traveller & & $\beta_{\text{female, cycle}}$ & $\beta_{\text{female, transit}}$ & $\beta_{\text{female, drive}}$ \\
\quad Traveller age $< 18$ years & & & $\beta_{\text{age} < 18 \text{ years, transit}}$ & $\beta_{\text{age} < 18 \text{ years, drive}}$ \\
\quad Traveller age $\geq 65$ years & & & $\beta_{\text{age} \geq 65 \text{ years, transit}}$ & $\beta_{\text{age} \geq 65 \text{ years, drive}}$ \\
\quad Travel during winter period (Nov--Mar) & & $\beta_{\text{winter, cycle}}$ \\
\quad No. of household cars & & & & $\beta_{\text{cars, drive}}$ \\
\bottomrule
\end{tabular}

%% file: table_cs_fit.tex
\begin{tabular}{l|c|cc|cc}
\toprule
{} &     \textbf{No. of} & \multicolumn{2}{c|}{\textbf{Log-lik.}} & \multicolumn{2}{c}{\textbf{AIC}} \\
\textbf{Model} & \textbf{parameters} &    \textbf{Train} &   \textbf{Test} &   \textbf{Train} &   \textbf{Test} \\
\midrule
MNP     &       23 &  -8297.2 & -960.9 & 16640.4 & 1967.9 \\
MNR     &       24 &  -8101.3 & -955.3 & 16250.5 & 1958.6 \\
Gen-MNR &       26 &  -8092.8 & -953.9 & 16237.6 & 1959.8 \\
\bottomrule
\end{tabular}

%% file: table_cs_results.tex
\begin{tabular}{l | rrrr | rrrr | rrrr}
\toprule
{} & \multicolumn{4}{c|}{\textbf{MNP}} & \multicolumn{4}{c|}{\textbf{MNR}} & \multicolumn{4}{c}{\textbf{Gen-MNR}} \\
\textbf{Parameter} &    \textbf{Mean} & \textbf{Std. dev.} & \textbf{[0.025\%} & \textbf{0.975\%]} &    \textbf{Mean} & \textbf{Std. dev.} & \textbf{[0.025\%} & \textbf{0.975\%]} &    \textbf{Mean} & \textbf{Std. dev.} & \textbf{[0.025\%} & \textbf{0.975\%]} \\
\midrule
$\beta_{\text{asc, cycle}}$                                        &  -2.168 &     0.126 &   -2.386 &   -1.892 &  -3.159 &     0.363 &   -3.762 &   -2.412 &  -1.839 &     0.140 &   -2.116 &   -1.540 \\
$\beta_{\text{asc, transit}}$                                      &  -0.290 &     0.028 &   -0.346 &   -0.238 &  -0.624 &     0.059 &   -0.733 &   -0.498 &  -0.634 &     0.056 &   -0.740 &   -0.532 \\
$\beta_{\text{asc, drive}}$                                        &  -0.819 &     0.056 &   -0.927 &   -0.707 &  -1.615 &     0.112 &   -1.822 &   -1.368 &  -1.738 &     0.092 &   -1.916 &   -1.563 \\
$\beta_{\text{cost}}$                                              &  -0.058 &     0.009 &   -0.075 &   -0.042 &  -0.112 &     0.018 &   -0.149 &   -0.077 &  -0.107 &     0.015 &   -0.137 &   -0.078 \\
$\beta_{\text{ovtt}}$                                              &  -1.878 &     0.121 &   -2.142 &   -1.645 &  -4.093 &     0.290 &   -4.666 &   -3.482 &  -3.874 &     0.235 &   -4.333 &   -3.472 \\
$\beta_{\text{ivtt}}$                                              &  -1.106 &     0.085 &   -1.272 &   -0.943 &  -2.457 &     0.206 &   -2.869 &   -2.025 &  -2.267 &     0.142 &   -2.548 &   -2.004 \\
$\beta_{\text{tv}}$                                                &  -1.281 &     0.090 &   -1.459 &   -1.102 &  -2.370 &     0.167 &   -2.696 &   -2.026 &  -2.248 &     0.110 &   -2.466 &   -2.041 \\
$\beta_{\text{transfers}}$                                         &  -0.081 &     0.016 &   -0.113 &   -0.049 &  -0.147 &     0.031 &   -0.208 &   -0.088 &  -0.149 &     0.029 &   -0.205 &   -0.092 \\
$\beta_{\text{female, cycle}}$                                     &  -0.456 &     0.076 &   -0.607 &   -0.312 &  -1.476 &     0.310 &   -2.134 &   -0.917 &  -1.249 &     0.425 &   -2.303 &   -0.722 \\
$\beta_{\text{winter, cycle}}$                                     &  -0.193 &     0.065 &   -0.324 &   -0.067 &  -0.436 &     0.171 &   -0.824 &   -0.136 &  -0.300 &     0.104 &   -0.542 &   -0.128 \\
$\beta_{\text{female, transit}}$                                   &   0.057 &     0.018 &    0.021 &    0.095 &   0.090 &     0.030 &    0.033 &    0.149 &   0.098 &     0.026 &    0.047 &    0.150 \\
$\beta_{\text{age} < 18 \text{ years, transit}}$                   &   0.096 &     0.031 &    0.038 &    0.159 &   0.116 &     0.048 &    0.024 &    0.213 &   0.116 &     0.044 &    0.030 &    0.202 \\
$\beta_{\text{age} \geq 65 \text{ years, transit}}$                &   0.177 &     0.027 &    0.125 &    0.231 &   0.286 &     0.045 &    0.200 &    0.374 &   0.261 &     0.042 &    0.181 &    0.344 \\
$\beta_{\text{female, drive}}$                                     &   0.055 &     0.026 &    0.007 &    0.107 &   0.020 &     0.046 &   -0.069 &    0.110 &   0.083 &     0.040 &    0.004 &    0.165 \\
$\beta_{\text{age} < 18 \text{ years, drive}}$                     &  -0.375 &     0.048 &   -0.472 &   -0.285 &  -0.795 &     0.086 &   -0.963 &   -0.627 &  -0.746 &     0.072 &   -0.891 &   -0.605 \\
$\beta_{\text{age} \geq 65 \text{ years, drive}}$                  &   0.177 &     0.036 &    0.108 &    0.247 &   0.305 &     0.067 &    0.174 &    0.440 &   0.287 &     0.062 &    0.164 &    0.409 \\
$\beta_{\text{cars, drive}}$                                       &   0.558 &     0.034 &    0.490 &    0.622 &   1.023 &     0.061 &    0.888 &    1.138 &   1.028 &     0.030 &    0.969 &    1.087 \\
\midrule
$\Sigma_{\text{walk}-\text{drive},\text{walk}-\text{drive}}$       &   0.698 &     0.085 &    0.524 &    0.858 &   1.287 &     0.102 &    1.026 &    1.451 &   1.619 &     0.083 &    1.474 &    1.796 \\
$\Sigma_{\text{walk}-\text{drive},\text{cycle}-\text{drive}}$      &   0.048 &     0.151 &   -0.198 &    0.375 &  -0.383 &     0.322 &   -0.855 &    0.224 &   0.473 &     0.096 &    0.264 &    0.636 \\
$\Sigma_{\text{walk}-\text{drive},\text{transit}-\text{drive}}$    &   0.445 &     0.053 &    0.338 &    0.546 &   0.948 &     0.079 &    0.731 &    1.084 &   1.336 &     0.034 &    1.255 &    1.390 \\
$\Sigma_{\text{cycle}-\text{drive},\text{cycle}-\text{drive}}$     &   1.938 &     0.124 &    1.713 &    2.185 &   0.922 &     0.166 &    0.631 &    1.363 &   0.221 &     0.069 &    0.102 &    0.373 \\
$\Sigma_{\text{cycle}-\text{drive},\text{transit}-\text{drive}}$   &   0.371 &     0.095 &    0.199 &    0.565 &  -0.169 &     0.287 &   -0.583 &    0.357 &   0.397 &     0.084 &    0.237 &    0.566 \\
$\Sigma_{\text{transit}-\text{drive},\text{transit}-\text{drive}}$ &   0.363 &     0.042 &    0.279 &    0.448 &   0.791 &     0.071 &    0.604 &    0.924 &   1.160 &     0.060 &    1.034 &    1.268 \\
\midrule
$\nu$                                                              &         &           &          &          &   2.132 &     0.209 &    1.793 &    2.658 &         &           &          &          \\
$\nu_{\text{walk}-\text{drive}}$                                   &         &           &          &          &         &           &          &          &   4.737 &     0.661 &    3.595 &    6.254 \\
$\nu_{\text{cycle}-\text{drive}}$                                  &         &           &          &          &         &           &          &          &   1.399 &     0.174 &    1.087 &    1.732 \\
$\nu_{\text{transit}-\text{drive}}$                                &         &           &          &          &         &           &          &          &  17.647 &     7.682 &    6.012 &   36.764 \\
\bottomrule
\end{tabular}

%% file: table_cs_wtp.tex
\begin{tabular}{ll|rrrr}
\toprule
\textbf{Attribute}                            & \textbf{Model}        &  \textbf{Mean} & \textbf{Std. dev.} & \textbf{[0.025\%} & \textbf{0.975\%]} \\
\midrule
Out-of-vehicle travel time [GBP/h] & MNP & 32.71 &      4.56 &    25.70 &    43.36 \\
                          & MNR & 37.22 &      5.56 &    28.80 &    50.93 \\
                          & Gen-MNR & 36.77 &      5.09 &    28.78 &    48.54 \\
\midrule                          
In-vehicle travel time [GBP/h] & MNP & 19.26 &      2.77 &    14.88 &    25.73 \\
                          & MNR & 22.33 &      3.41 &    17.02 &    30.44 \\
                          & Gen-MNR & 21.52 &      3.04 &    16.63 &    28.44 \\
\midrule
Transfers [GBP/interchange] & MNP &  1.42 &      0.40 &     0.76 &     2.34 \\
                          & MNR &  1.36 &      0.42 &     0.67 &     2.36 \\
                          & Gen-MNR &  1.44 &      0.40 &     0.75 &     2.34 \\
\midrule
Traffic variability [GBP] & MNP & 22.34 &      3.42 &    17.00 &    30.48 \\
                          & MNR & 21.60 &      3.57 &    16.11 &    30.14 \\
                          & Gen-MNR & 21.40 &      3.26 &    16.24 &    28.81 \\
\bottomrule
\end{tabular}

%% file: table_cs_elas.tex
\begin{tabular}{ll|l|rrrr}
\toprule
\textbf{Scenario}                                   &                   & \textbf{Model}          &  \textbf{Walk} &  \textbf{Cycle} &  \textbf{Transit} &  \textbf{Drive} \\
\midrule
Cycling out-of-vehicle travel time & decreased by 10\% & MNP & -0.00 &  -0.94 &     0.05 &   0.02 \\
                                   &                   & MNR &  0.00 &  -0.76 &     0.02 &   0.03 \\
                                   &                   & Gen-MNR &  0.01 &  -1.03 &     0.05 &   0.02 \\
\midrule
Walking out-of-vehicle travel time & decreased by 10\% & MNP & -1.57 &   0.02 &     0.52 &   0.16 \\
                                   &                   & MNR & -1.78 &   0.04 &     0.58 &   0.16 \\
                                   &                   & Gen-MNR & -1.78 &   0.27 &     0.57 &   0.16 \\
\midrule
Driving traffic variability & decreased by 10\% & MNP &  0.13 &   0.17 &     0.38 &  -0.40 \\
                                   &                   & MNR &  0.14 &   0.31 &     0.35 &  -0.38 \\
                                   &                   & Gen-MNR &  0.13 &   0.20 &     0.35 &  -0.38 \\
\midrule
Transit in-vehicle travel time & decreased by 10\% & MNP &  0.18 &   0.27 &    -0.40 &   0.26 \\
                                   &                   & MNR &  0.21 &   0.11 &    -0.43 &   0.29 \\
                                   &                   & Gen-MNR &  0.20 &   0.28 &    -0.43 &   0.29 \\
\midrule                                   
Transit out-of-vehicle travel time & decreased by 10\% & MNP &  0.34 &   0.31 &    -0.45 &   0.23 \\
                                   &                   & MNR &  0.43 &   0.12 &    -0.49 &   0.26 \\
                                   &                   & Gen-MNR &  0.42 &   0.31 &    -0.51 &   0.26 \\
\bottomrule
\end{tabular}